\documentclass[aps,onecolumn,pra,superscriptaddress]{revtex4}
\usepackage{amsmath,amssymb,amsfonts,bbm,graphicx,times,psfrag}
\usepackage[pdftex]{color}
\usepackage{lineno}
\usepackage[dvipsnames]{xcolor}
\usepackage{amsmath,bm}
\usepackage[colorlinks, linkcolor=blue, citecolor=blue, urlcolor=blue, breaklinks=true]{hyperref}
\usepackage{microtype}

\usepackage{amsthm}
\usepackage{csquotes}
\usepackage{amsthm}\newcommand{\be} {\begin{equation}}
\newcommand{\ee} {\end{equation}}

\newcommand{\Tr}{{\rm Tr}}

\usepackage{color}
\usepackage{mathrsfs}

\newcommand{\ic}{{i}}
\newcommand{\e}{{e}}

\begin{document}

\title{Nonreciprocal photon blockade in a spinning microwave magnomechanical system through Kerr-magnon and optical parametric amplifier}

\author{S. K. Singh} 
\affiliation{Department of Physics, Akal University, Talwandi Sabo, Bathinda, Punjab 151302, India}
\author{Mohamed Amazioug} \thanks{m.amazioug@uiz.ac.ma}
\affiliation{LPTHE, Department of Physics, Faculty of sciences, Ibnou Zohr University, Agadir, Morocco}
\author{Jia-Xin Peng} 
\affiliation{School of Physical Science and Technology, Nantong University, Nantong 226019, People’s Republic of China}
\author{Wedad Albalawi} 
\affiliation{Department of Mathematical Sciences, College of Science, Princess Nourah bint Abdulrahman University, P.O. Box 84428, Riyadh 11671, Saudi Arabia} 
\author{Mohammad Khalid}
\affiliation{Sunway Centre for Electrochemical Energy and Sustainable Technology (SCEEST), Faculty of Engineering and Technology, Sunway University, No. 5 Jalan Universiti, Bandar Sunway, 47500 Petaling Jaya, Selangor, Malaysia}
\affiliation{University Centre for Research and Development, Chandigarh University, Mohali, Punjab, 140413, India}
\author{Abdel-Haleem Abdel-Aty} 
\affiliation{Department of Physics, College of Sciences, University of Bisha, Bisha 61922, Saudi Arabia}

\begin{abstract}

Unconventional quantum antibunching, arising from quantum interference effects, represents a notable form of quantum correlation that has attracted significant attention for its ability to generate high-quality single-quantum sources. In this work, we propose a scheme to achieve and actively control strong photon blockade in a spinning microwave magnomechanical system by leveraging the combined nonlinear effects of Kerr-induced magnon interactions and an optical parametric amplifier. By exploiting the Sagnac-Fizeau shift, we establish nonreciprocal photon blockade and verify this effect through a combination of analytical modelling and numerical simulations. To gain intuitive insight into the underlying nonreciprocity, we approximate the equal-time second-order correlation function using the analytical solution of the Schr\"odinger equation. This analytical result is then compared with the full numerical solution derived from the Lindblad master equation. The influences of thermal noise, the probe field amplitude, and the magnetic-dipole coupling strength are investigated within the constraints of the weak-coupling regime. The system's nonclassicality is characterized using the Mandel parameter, complemented by an analysis of the time evolution of the second-order correlation function. Our work provides a pathway for realizing nonreciprocal photon blockade in a nonlinear spinning microwave magnomechanical system.

\end{abstract} 
\maketitle 

\section{Introduction}

The photon blockade (PB) effect, a key mechanism for single-photon generation, arises when the presence of one photon in a nonlinear medium prevents the entry of another. It plays a pivotal role in quantum metrology ~\cite{Giovannetti2011} and quantum information processing~\cite{Stannigel2012,Bennett2000,Buluta2011}, enabling precise measurements and reliable single-photon sources for quantum communication and computing. Photon blockade (PB) is mainly classified into two categories. The first is conventional photon blockade (CPB), which originates from the anharmonicity of energy levels and requires strong nonlinear interactions to suppress multi-photon excitations. The second is unconventional photon blockade (UPB), which relies on destructive quantum interference between different excitation pathways, allowing photon blockade to occur even in systems with weak nonlinearities. Experimental realizations of CPB have been reported in systems like photonic crystals with quantum dots~\cite{Faraon2008}, cavity QED~\cite{Birnbaum2005,Reinhard2012,Muller2015,Hamsen2017,Zheng2023}, and circuit QED~\cite{Lang2011,Hoffman2011}. In contrast, UPB has been demonstrated in circuit QED~\cite{Vaneph2018} and coupled quantum-dot-cavity systems~\cite{Snijders2018}.\\

Optical reciprocity requires a system to exhibit exchange symmetry, meaning the transmission response remains the same when the input and output ports are interchanged. Optical nonreciprocal devices, such as optical isolators ~\cite{Sayrin2015,Tang2019} and circulators ~\cite{Kamal2011}, break this symmetry, allowing light to travel in only one direction. This one-way behavior helps protect light sources from unwanted back reflections and noise, making these devices vital for applications like invisible sensing ~\cite{Sounas2017} and backaction-immune optical communications ~\cite{Svela2020} as well as  enabled effects like nonreciprocal entanglement~\cite{Jiao2020,Jiao2022,Ren2022,Chen2023}, phonon lasing~\cite{Jiang2018,Xu2021}, slow light~\cite{Mirza2019,Peng2023}, and optical solitons~\cite{Li2021}. In 2018, nonreciprocal PB was theoretically proposed in a spinning Kerr resonator~\cite{Huang2018}, followed by its realization in various platforms which includes nonlinear cavities~\cite{Wang2019,Shen2020,Xu2020}, optomechanical systems~\cite{Shang2021,Liu2023,Li2019}, atom-cavity systems~\cite{Liu2023a,Zhang2023,Xue2020,Jing2021}, and cavity optomagnonics~\cite{Xie2022}. Experimental implementations based on cavity optomechanical systems  face experimental challenges as because of requirement of maintaining a stable inter-cavity coupling.\\

Spining microwave magnomechanical systems may offer a compelling alternative as such  systems benefit from strong photon-magnon coupling via Kittel modes in YIG spheres~\cite{Kittel1948,Zhang2014,Huebl2013,Bai2015,Tabuchi14} and tunable magnon frequencies via external magnetic fields. Moreover, the magnetostrictive interaction provides controllable coupling with phonons~\cite{Kippenberg2005,Zhao2020,Zhang16,Li18,Li20},  and superconducting qubits \cite{Tabuchi15,Lachance20}. Nonreciprocal quantum phenomena such as entanglement~\cite{Yang2020} and phonon lasing~\cite{Xu2021} have recently been explored in these platforms. These developments motivate the study of nonreciprocal UPB in spinning microwave magnomechanical systems. With the  help of well known Sagnac effect and quantum interference, it might also becomes possible to achieve direction-dependent photon statistics, laying the foundation for tunable nonreciprocal single-photon sources in cavity magnomechanical systems. Consequently, magnomechanics presents an exciting platform for exploring various quantum phenomena, such as quantum correlations \cite{FB23,FBEntropy,FB-EPJP,SR25}, cooling \cite{Asjad23}, non-reciprocity (including photon and magnon blockade) \cite{Huang18,mPB,Xu24,Hou25,Ge25,Zhang25}, and thermodynamics \cite{Amghar24,Edet24,DFDB25}.\\

With advancements in the quality and performance of nonlinear optical crystals, optical parametric amplifiers (OPAs), including both degenerate and nondegenerate configurations, have become essential tools for generating high-purity squeezed and entangled quantum states \cite{Collett84,Agarwal06,He07,Yan12}. Interference effects in the quantum fluctuations of OPA outputs were analysed, and it was shown that such interference phenomena can be precisely controlled by tailoring the squeezing properties of the input field \cite{Agarwal06}. The generation of broadband entangled light via cascaded nondegenerate OPAs was reported, and its applicability to broadband quantum teleportation was discussed, demonstrating strong potential for continuous-variable quantum information processing \cite{He07}. More recently, enhancement of entanglement has been investigated through the implementation of nondegenerate OPAs \cite{Chen09,Shang10,Zhou15}. It was theoretically demonstrated that the correlation strength of pre-existing entangled beams can be significantly improved by incorporating a nondegenerate OPA inside an optical cavity \cite{Chen09}, while experimental studies verified that both the degree and tunability of entanglement can be effectively enhanced using nondegenerate OPA-based architectures \cite{Shang10,Zhou15}, thereby providing a promising pathway for advanced quantum resource engineering in photonic platforms.\\

It is important to clarify the different physical roles played by Kerr nonlinearity and the optical parametric amplifier (OPA) in our proposed scheme. The Kerr-type magnon nonlinearity originates from intrinsic magnetocrystalline anisotropy and higher-order spin interactions in the YIG sphere which will give an effective term proportional to $\rm m^{\dagger} m^{\dagger} m m$ in our system Hamiltonian. This nonlinear interaction results the magnon energy spectrum anharmonic, such that the two-excitation manifold acquires an additional energy shift as compared to the single-excitation state. As a result, the effective detuning for multiphoton transitions is shifted. The competing excitation pathways depend strongly on this detuning, so the Kerr interaction enables controlled adjustment of their relative phase. This phase control is essential for realizing the destructive interference required for unconventional photon blockade. In contrast, the OPA provides a direct two-photon excitation pathways that coherently couples the vacuum state to the two-photon state. The coexistence of sequential excitation pathways and OPA-induced parametric pathways enables destructive quantum interference even in the weak-nonlinearity regime. When it is combined with the rotation-induced Sagnac–Fizeau shift, which breaks the symmetry of the effective detuning for clockwise and counterclockwise propagation, this interference mechanism becomes direction dependent, giving rise to nonreciprocal unconventional photon blockade. In this work, we theoretically propose  a scheme for realizing nonreciprocal unconventional photon blockade in a spinning microwave magnomechanical system incorporating Kerr-magnon interaction and a degenerate optical parametric amplifier under weak driving conditions. Strong photon antibunching is achieved through precise control of the nonlinear parameters, as confirmed by both analytical derivations and numerical simulations using experimentally feasible parameters. We onwards explore the influence of thermal noise, probe-field amplitude, and magnetic-dipole coupling strength on the robustness of the blockade, and characterize the nonclassical properties of the system using the Mandel $Q$ parameter and second-order correlation functions.

As Compared to previously proposed nonreciprocal photon blockade schemes based on chiral waveguides, cascaded cavities, parity-time symmetric systems, or purely Kerr-type nonlinear interactions our present proposal exhibits number of distinctive advantages. The main advantage of our scheme is that the nonreciprocity here does not rely on asymmetric dissipation or synthetic gauge fields, but instead originates from rotation-induced Sagnac-Fizeau detuning asymmetry which provides a controllable and physically transparent mechanism. Second, unconventional photon blockade is achieved even under weak Kerr nonlinearity through interference engineering assisted by an optical parametric amplifier, significantly relaxing the requirement of strong intrinsic nonlinearity. Third, this cavity magnomechanical platform offers tunable magnetic-dipole coupling and mechanical degrees of freedom, enabling flexible parameter control that is not readily available in purely optical systems. These features collectively distinguish our scheme from existing approaches and highlight its potential for integrated and direction-dependent quantum photonic devices.

The remainder of this paper is organized as follows: Section II describes the spinning microwave magnomechanical system-incorporating both a Kerr-magnon and an optical parametric amplifier-and provides the system Hamiltonian in the weak-coupling regime. In Section III, we employ the second-order correlation function to characterize the nonreciprocal photon blockade under optimal parameters, utilizing both analytical and numerical approaches. Section IV presents the results and provides a detailed discussion of our findings. Finally, Section V concludes the paper.

\begin{figure}[!htb]
\begin{center}
\includegraphics[width=1\columnwidth]{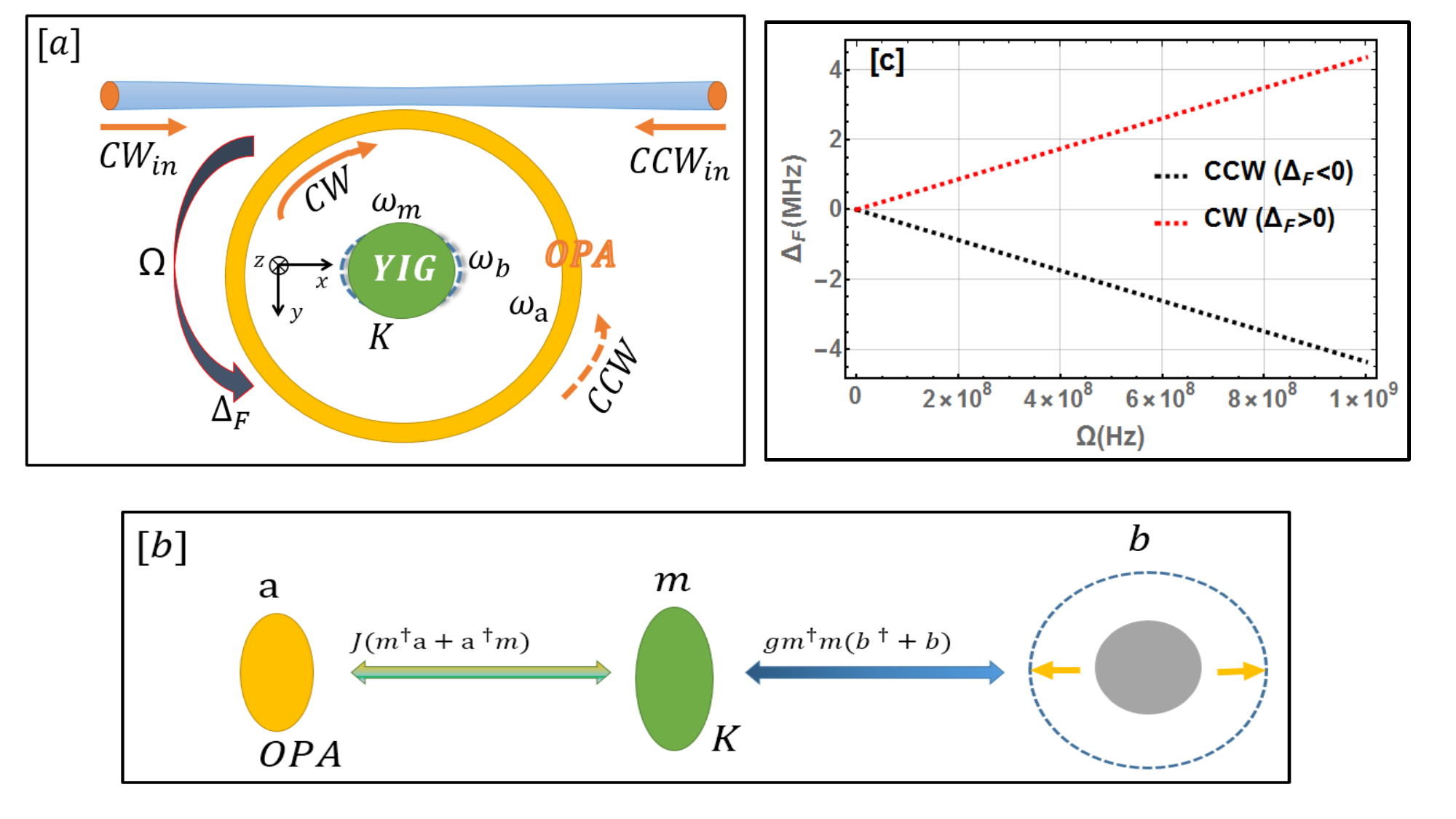}
\end{center}
\caption{[a] The panel illustrates a spinning microwave magnetomechanical system, incorporating magnon squeezing. A spinning optical resonator (mode $\rm a$) is driven from either the left or the right side, generating a Sagnac-Fizeau shift ($\Delta_F$), where the sign corresponds to the drive direction ($\Delta_F \gtrless 0$). A bias magnetic field ${\rm H}$ fully magnetizes the YIG sphere, which supports the magnon mode ($m$) and the phonon mode ($b$). Crucially, the YIG sphere's rotation at angular frequency $\Delta_B$ induces an emergent magnetic field $\mathbf{H}$, which shifts the magnon mode frequency. The system utilizes an optical parametric amplifier (OPA). In this process, a single pump photon at frequency $\omega_{0}$ is down-converted into a pair of identical cavity photons (signal and idler) at frequency $\omega_{\rm c}$, satisfying the resonance condition $\omega_0 = 2\omega_{\rm c}$. [b] photons in the cavity couple to the magnons via the magnetic dipole interaction, while magnons couple to the phonons through the magnetostrictive effect, which functions similarly to radiation pressure. [c] Plot of Sagnac-Fizeau shift $\Delta_F$ as a function of angular velocity $\Omega$.}
\label{NThm1}  
\end{figure}

\section{Model Hamiltonian}

As shown in Fig. \ref{NThm1}, we investigate a spinning cavity magnomechanical system. The resonant frequencies of its counter-propagating modes experience a Sagnac-Fizeau shift given by \cite{Malykin2000}

\begin{equation} \label{1}
\Delta_F = \pm \Omega \frac{{\rm nr} \omega_a}{c} \left(1 - \frac{1}{n^2} - \frac{\lambda}{n} \frac{dn}{d\lambda} \right),
\end{equation}
Here \( \Omega \) is the angular velocity, \( {\rm n} \) and \( {\rm r} \) are the refractive index and radius of the microwave resonator, \( c \) is the speed of light in vacuum, \( \lambda \) is the wavelength, and \( \omega_a \) is the resonant frequency of the nonspinning cavity. Following \cite{Maayani18}, we neglect the dispersion term $\dfrac{\mathrm{d} {\rm n}}{\mathrm{d} \lambda}$ as its relativistic contribution to the Sagnac effect is small ($\sim 1\%$). In Eq. (\ref{1}), the signs $+$ and $-$ correspond to the clockwise and counterclockwise driving fields, respectively, under the assumption of a clockwise-rotating cavity. The sign of $\Delta_F$ determines the direction of the driving field: $\Delta_F > 0$ corresponds to a clockwise (CW) field, while $\Delta_F < 0$ denotes a counterclockwise (CCW) field, as shown in Fig. \ref{NThm1}(c). Consequently, the shifted frequencies are given by $\omega_{\text{CW, CCW}} = \omega_a \pm |\Delta_F|$. In a frame rotating at the drive frequency $\omega_d$, defined by the unitary operator $U(t) = e^{-i\omega_d(\rm a^\dagger a + m^\dagger m)t}$, the time-independent Hamiltonian of the system becomes
\begin{equation} \label{Eq2}
\mathscr{H}_1 = \mathscr{H}_0 + \mathscr{H}_I + \mathscr{H}_K + \mathscr{H}_{OPA}+ \mathscr{H}_d,
\end{equation}
where the free Hamiltonian is
\begin{equation}
\mathscr{H}_0 = (\Delta_{\rm a} + \Delta_F) {\rm a}^\dagger {\rm a} + \Delta_{\rm m} {\rm m}^\dagger {\rm m} + \omega_b b^\dagger b,
\end{equation}
with \( {\rm a}^\dagger ({\rm a}), {\rm m}^\dagger ({\rm m}), b^\dagger (b) \) being the creation (annihilation) operators for microwave cavity, magnon, and phonon modes, respectively. \( \Delta_{\rm a} = \omega_a - \omega_d \) and \( \Delta_{\rm m} = \omega_m - \omega_d \). The interaction Hamiltonian is
\begin{equation}
\mathscr{H}_I = J({\rm a}^\dagger {\rm m} + {\rm a} {\rm m}^\dagger) - g {\rm m}^\dagger {\rm m} (b^\dagger + b),
\end{equation}
where \( J \) is the magnon-photon coupling strength and \( g \) is the magnon-phonon coupling strength. The Hamiltonian of Kerr, writes as
\begin{equation}
\mathscr{H}_K = K_0 ({\rm m}^\dagger {\rm m})^2,
\end{equation}

and Hamiltonian of the parametric amplifier

\begin{equation}
\mathscr{H}_{OPA} = i\Lambda [({\rm a}^\dagger)^2\e^{i\beta}-{\rm a}^2\e^{-i\beta}],
\end{equation}
where $\Lambda$ is the squeezing parameter, quantifying the degree of squeezing, and $\beta$ is the squeezing angle (phase). The driving term is given by
\begin{equation}
\mathscr{H}_d = \epsilon({\rm a}^\dagger + {\rm a}),
\end{equation}
with \( \varepsilon \) being the driving amplitude. In the limit of weak magnomechanical coupling ($g/\omega_b \ll 1$), we apply a unitary displacement transformation $\mathrm{U} = \exp[g(b - b^\dagger)/\omega_b]$ to the Hamiltonian $\mathscr{H}_1$. This transformation effectively decouples the mechanical and magnonic degrees of freedom, the effective Hamiltonian becomes
\begin{equation}
\mathscr{H}_2 = (\Delta_{\rm a} + \Delta_F) {\rm a}^\dagger {\rm a} + \Delta_{\rm m} {\rm m}^\dagger {\rm m} + \omega_b b^\dagger b + K ({\rm m}^\dagger {\rm m})^2 + J({\rm a}^\dagger {\rm m} + {\rm a} {\rm m}^\dagger)+i\Lambda [({\rm a}^\dagger)^2\e^{i\beta}-{\rm a}^2\e^{-i\beta}] + \epsilon({\rm a}^\dagger + {\rm a}),
\end{equation}
where the Kerr-type nonlinear strength became \( K = K_0 - \frac{g^2}{\omega_b} \). Neglecting the decoupled phonon term, the reduced Hamiltonian is
\begin{equation} \label{9}
\mathscr{H}'_2 = (\Delta_{\rm a} + \Delta_F) {\rm a}^\dagger {\rm a} + \Delta_{\rm m} {\rm m}^\dagger {\rm m} + K ({\rm m}^\dagger {\rm m})^2 + J({\rm a}^\dagger {\rm m} + {\rm a} {\rm m}^\dagger)+i\Lambda [({\rm a}^\dagger)^2\e^{i\beta}-{\rm a}^2\e^{-i\beta}] + \epsilon({\rm a}^\dagger + {\rm a}).
\end{equation}

By including phenomenological attenuation, the effective non-Hermitian Hamiltonian is expressed as
\begin{equation}
\mathscr{H} = \bigg(\Delta_{\rm a} + \Delta_F -\ic\frac{\gamma_{\rm a}}{2}\bigg) {\rm a}^\dagger {\rm a} + \bigg(\Delta_{\rm m}-\ic\frac{\gamma_{\rm m}}{2}\bigg) {\rm m}^\dagger {\rm m} + K ({\rm m}^\dagger {\rm m})^2 + J({\rm a}^\dagger {\rm m} + {\rm a} {\rm m}^\dagger) +i\Lambda [({\rm a}^\dagger)^2\e^{i\beta}-{\rm a}^2\e^{-i\beta}] + \epsilon({\rm a}^\dagger + {\rm a}),
\end{equation}
with dissipation rates $\gamma_{\rm a}$ and $\gamma_{\rm m}$ for the cavity and Kittel modes, respectively.

\section{Nonreciprocal photon blockade} 

In this section, we focus on both the analytical solution of the non-Hermitian Schr\"odinger equation and the numerical results obtained via the master equation approach. Additionally, we investigate the feasibility of achieving a strong photon blockade in the weak-coupling limit ($g \ll \omega_b$). The analytical expression for the correlation function is obtained by solving the Schr\"odinger equation $i\partial_t |\Psi(t)\rangle = \mathscr{H}|\Psi(t)\rangle$ under the assumption of a weak driving field. This solution is found by truncating the Hilbert space to the low-excitation subspace $\mathscr{D}=\left\{ |{\rm d,f}\rangle \mid {\rm d+f}\leq 2 \right\}$, encompassing states with up to two total excitations. In the bare-state basis $|\rm d,f\rangle$ (where ${\rm d}$ and ${\rm f}$ are the magnon and photon numbers in the cavity), the system state within the ${\rm d+f}\leq2$ low-excitation subspace is written as 
\begin{equation} \label{eq:7} 
|\Psi(t)\rangle=\sum_{\rm d,f}^{{\rm d+f}\leq 2}\mathrm{C}_{\rm df}(t)|{\rm d, f}\rangle,
\end{equation}
where ${\rm C}_{\rm df}(t)$ denotes the time-dependent amplitude of the state $|{\rm d, f}\rangle$. The probability amplitudes $\mathrm{C}_{\rm df} (t)$ satisfy a set of linear differential equations derived from the Schr\"odinger equation %
\begin{equation} \label{eq:7} 
\ic\partial_t \mathrm{C}_{00} = \epsilon\mathrm{C}_{01}-i\sqrt{2}\Lambda\e^{-i\beta}\mathrm{C}_{02},
\end{equation}
\begin{equation} \label{eq:9} 
\ic\partial_t \mathrm{C}_{10} = \epsilon\mathrm{C}_{11}+(\Delta'_{\rm m} + K)\mathrm{C}_{10}+J\mathrm{C}_{01},
\end{equation}
\begin{equation} \label{eq:8} 
\ic\partial_t \mathrm{C}_{01} = J\mathrm{C}_{10}+(\Delta'_{\rm a}+\Delta_{F})\mathrm{C}_{01}+\epsilon\mathrm{C}_{00}+\sqrt{2}\epsilon\mathrm{C}_{02},
\end{equation}
\begin{equation} \label{eq:10} 
\ic\partial_t \mathrm{C}_{11}= \sqrt{2}J\mathrm{C}_{20}+\epsilon\mathrm{C}_{10}+(\Delta'_{\rm a} + \Delta_{F} + \Delta'_{\rm m} + K)\mathrm{C}_{11}+\sqrt{2}J\mathrm{C}_{02},
\end{equation}
\begin{equation} \label{eq:11} 
\ic\partial_t \mathrm{C}_{02}= \sqrt{2}J\mathrm{C}_{11}+2(\Delta'_{\rm a} + \Delta_{F})\mathrm{C}_{02}+\sqrt{2}\epsilon\mathrm{C}_{01}+i\sqrt{2}\Lambda\e^{i\beta}\mathrm{C}_{00},
\end{equation}
\begin{equation} \label{eq:12} 
\ic\partial_t \mathrm{C}_{20}= 2(\Delta'_{\rm m} + 2K)\mathrm{C}_{20}+\sqrt{2}J\mathrm{C}_{11},
\end{equation}
where $\Delta'_{\rm a}=\Delta_{\rm a}-\ic\gamma_{\rm a}/2$ and $\Delta'_{\rm m}=\Delta_{\rm m}-\ic\gamma_{\rm m}/2$. The equations (\ref{eq:7})-(\ref{eq:12}) are analytically solvable, yielding the dynamical state. For the steady state, we solve $\partial_t \mathrm{C}_{\rm df}=0$. In the case of weak coupling interaction, this equation is simplified using appropriate approximations, for instance, by ignoring higher-order terms. For simplicity, we set, $\Delta_{\rm a}=\Delta_{\rm m}=\Delta$ and $\gamma_{\rm a}=\gamma_{\rm m}=\gamma$. In the weak driving limit ($\epsilon\ll\gamma$), the probability amplitudes satisfy the hierarchy $|{\rm C}_{00}|\simeq1\gg|{\rm C}_{01}|,|{\rm C}_{10}|\gg|{\rm C}_{11}|,|{\rm C}_{02}|,|{\rm C}_{20}|$. The explicit expression of $\mathrm{C}_{01}$ is given by
\begin{equation} \label{eq:14} 
\mathrm{C}_{01} = -\frac{(K+\Delta'_{\rm m}) \epsilon }{-J^2+K \Delta'_{\rm a}+\Delta'_{\rm a}\Delta'_{\rm m}+K \Delta_{\rm F}+\Delta'_{\rm m} \Delta_{\rm F}},%
\end{equation}
and the explicit expression of $\mathrm{C}_{02}$ is given by 
\begin{equation} \label{eq:14} 
\mathrm{C}_{02} = A/(B_1\times B_2),
\end{equation}
where $A=J \left(J^2-2 K^2-4 K\Delta'_{\rm a}-3 K\Delta'_{\rm m}-2 \Delta'_{\rm a}\Delta'_{\rm m}-(\Delta'_{\rm m})^2-4 K \Delta_F-2\Delta'_{\rm m}\Delta_F\right) \epsilon ^2,$~~
$B_1= \sqrt{2} \left(J^2-(K+\Delta'_{\rm m}) (\Delta'_{\rm a}+\Delta_F)\right),~ and ~$
$B_2 =-(2 K+\Delta'_{\rm m}) (\Delta'_{\rm a}+\Delta_F) (K+\Delta'_{\rm a}+\Delta'_{\rm m}+\Delta_F)+J^2 (2 K+\Delta'_{\rm a}+\Delta'_{\rm m}+\Delta_F).$
The Photon blockade is quantified by the steady-state second-order correlation function $g_{\rm a}^{(2)}(0)$, given by 
\begin{equation} \label{eq:18}
g_{\rm a}^{(2)}(0)=\frac{2|\mathrm{C}_{02}|^2}{|\mathrm{C}_{01}|^4}=\frac{\Tr[{\rm a^{\dag} a^{\dag} a a}\varrho_{\rm s}]}{\Tr[{\rm a^{\dag} \rm a}\varrho_{\rm s}]}.
\end{equation}
The statistical properties of photon is characterized by the equal-time second-order correlation function $g_{\rm a}^{(2)}(0)$. Specifically, $g_{\rm a}^{(2)}(0)>1$ signifies bunching, while $g_{\rm a}^{(2)}(0)<1$ indicates antibunching. Perfect photon blockade occurs when $g_{\rm a}^{(2)}(0)=0$, with $g_{\rm a}^{(2)}(0)=0$ corresponds to strong photon blockade. This property is numerically investigated by solving the Master equation, which has the following expression
\begin{equation} \label{eq:21} 
\partial_t \varrho=-\ic [\mathscr{H}^\prime_2,\varrho]+\frac{\gamma}{2}\mathrm{L}_{\rm a}(\varrho)+\frac{\gamma}{2}[\rm m_{th}+1]\mathrm{L}_{\rm m}(\varrho)+\frac{\gamma}{2}\rm m_{th}\mathrm{L}_{\rm m^\dag}(\varrho)
\end{equation}
where $\varrho$ denotes the system's density matrix and ${\rm m}_{\rm th} = [\exp(\hbar\omega_m / k_B T) - 1]^{-1}$ represents the thermal equilibrium occupancy of the magnon mode at ambient temperature $T$ ($k_B$ being the Boltzmann constant), with $\mathscr{H}^\prime_2$ is the Hamiltonian from Equation (\ref{9}); $\mathrm{L}_{\rm a}$ and $\mathrm{L}_{\rm m}$ are the Lindblad superoperators given by $\mathrm{L}_{\rm a}(\varrho)=2{\rm a}\varrho {\rm a}^\dag-\{{\rm a}^{\dagger}{\rm a},\varrho\}$ and $\mathrm{L}_{\rm m}(\varrho)=2{\rm m}\varrho {\rm m}^\dag -\{{\rm m}^{\dagger}{\rm m},\varrho\}$ for the optical $\rm a$ and magnon $\rm m$ modes, respectively; and $\gamma$ is the decay rate of the  magnon (photon) mode. In steady state, the density matrix $\varrho$ is characterized by $\partial_t \varrho = 0$.

\section{RESULTS AND DISCUSSIONS}

The achieved photon blockade effect belongs to the UCPB (destructive quantum interference). Here we employ the parameters value considered \cite{Deng24}: $\gamma / 2\pi = 0.55 \times 10^{6} \text{ Hz}$, $\omega_a / 2\pi = 10.1 \times 10^{9} \text{ Hz}$, $\omega_b / 2\pi = 11.0308 \times 10^{6} \text{ Hz}$, $\epsilon = 0.005\gamma$, $J / 2\pi = 7.37 \times 10^{6} \text{ Hz}$, and $K=0.1\gamma$ \cite{Ebrahimi23}. The Sagnac-Fizeau shift $\Delta_F$ as a function angular velocity $\Omega$, with the refractive index ${\rm n}=1.4$, the resonator's radius ${\rm r}=30\mu$m, the vacuum wavelength of light $\lambda=1550n$m, and the vacuum speed of light $c=3\times 10^8$m/s \cite{Hou2025}. The optimal parameter pairs $\lambda$ and $\Delta$ can be obtained together with the other fixed parameters. Beside, the real solution of the optimal parameter pairs value:\\
* $\Delta_F=  0.5\gamma$  :  for photon blockade $\{ \Delta_{opt} = -0.684495 \omega_b, \Lambda_{opt} = 2.46157\times 10^{-6}\omega_b \};\{ \Delta_{opt} =  0.654639 \omega_b, \Lambda_{opt} = 2.45563\times 10^{-6}\omega_b \}$.\\
* $\Delta_F= - 0.5\gamma$  :  for photon blockade $\{ \Delta_{opt} = 0.679535 \omega_b, \Lambda_{opt} = 2.46105\times 10^{-6}\omega_b \};\{ \Delta_{opt} =  -0.659796 \omega_b, \Lambda_{opt} = 2.47275\times 10^{-6}\omega_b \}$.
\begin{figure}[!htb]
\begin{center}
\includegraphics[width=8cm,height=6cm]{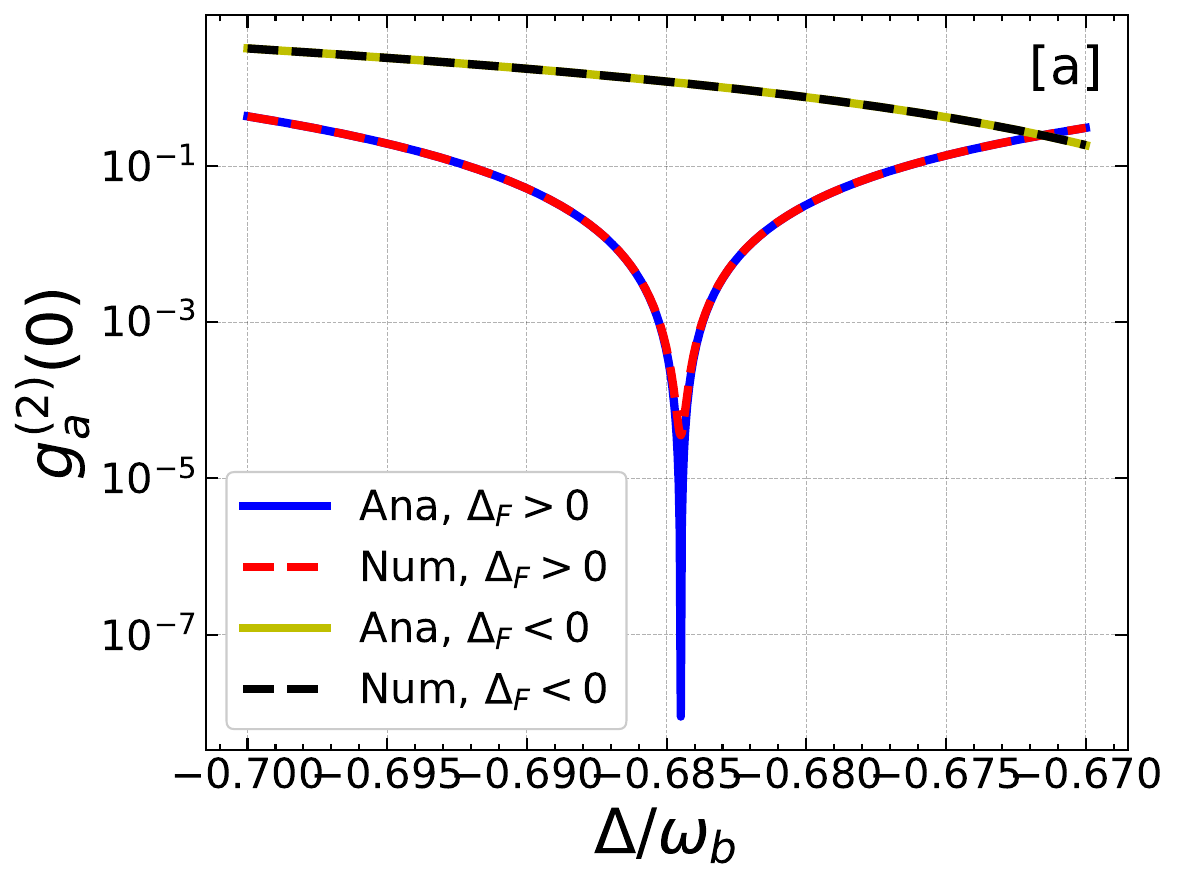}
\includegraphics[width=8cm,height=6cm]{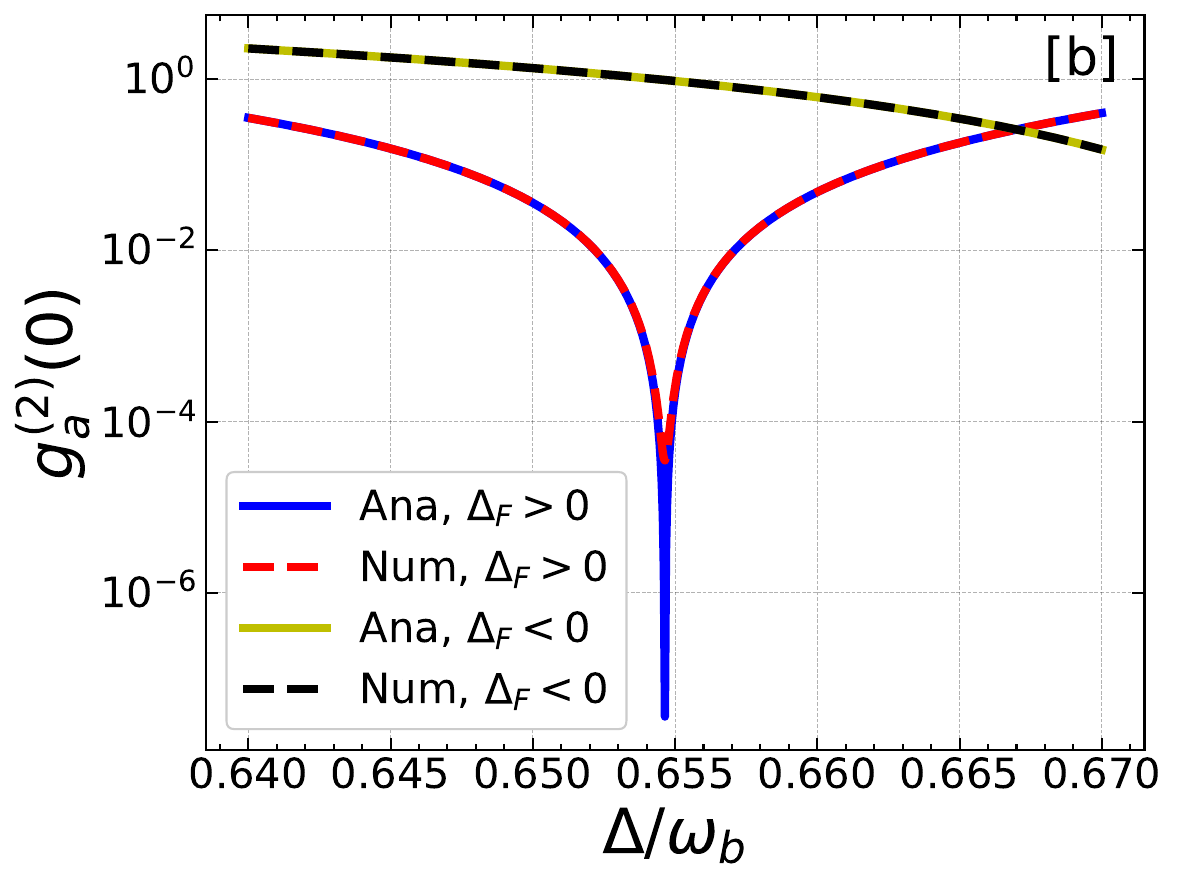}
\includegraphics[width=8cm,height=6cm]{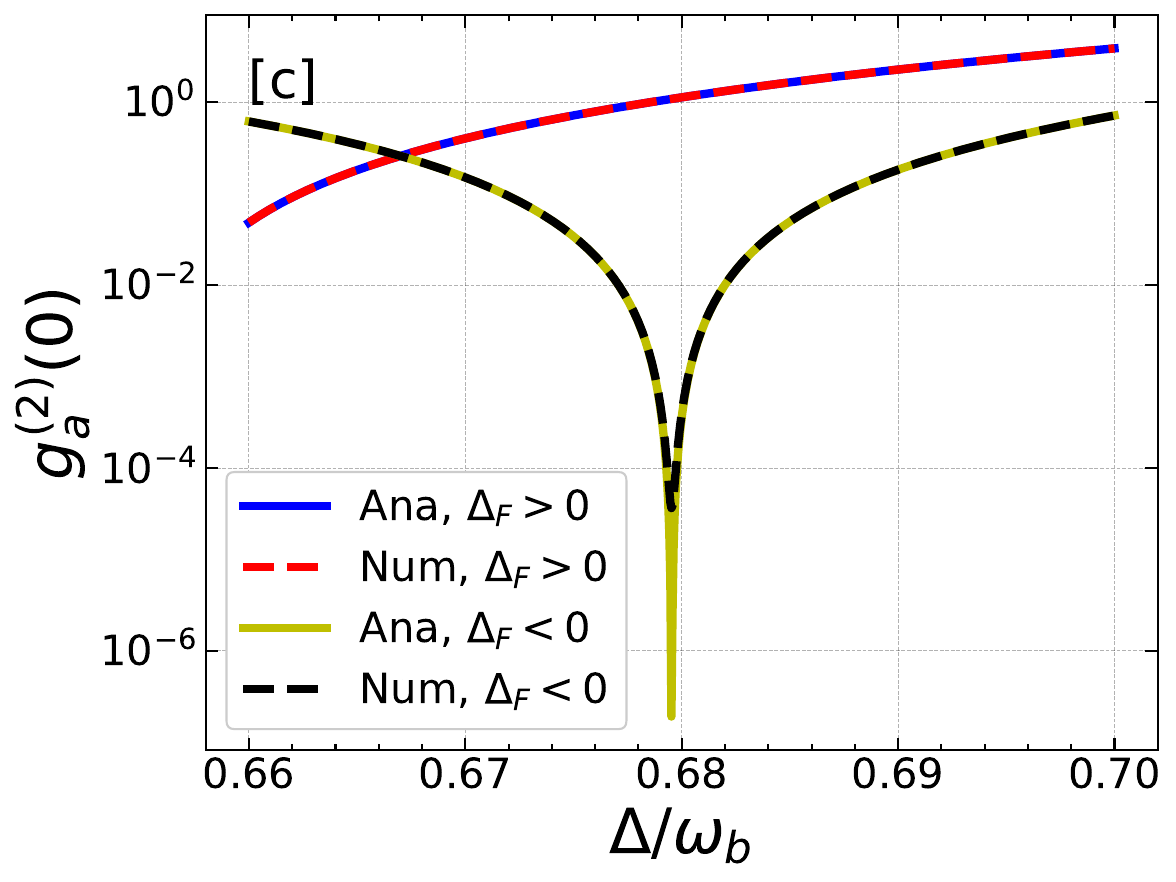}
\includegraphics[width=8cm,height=6cm]{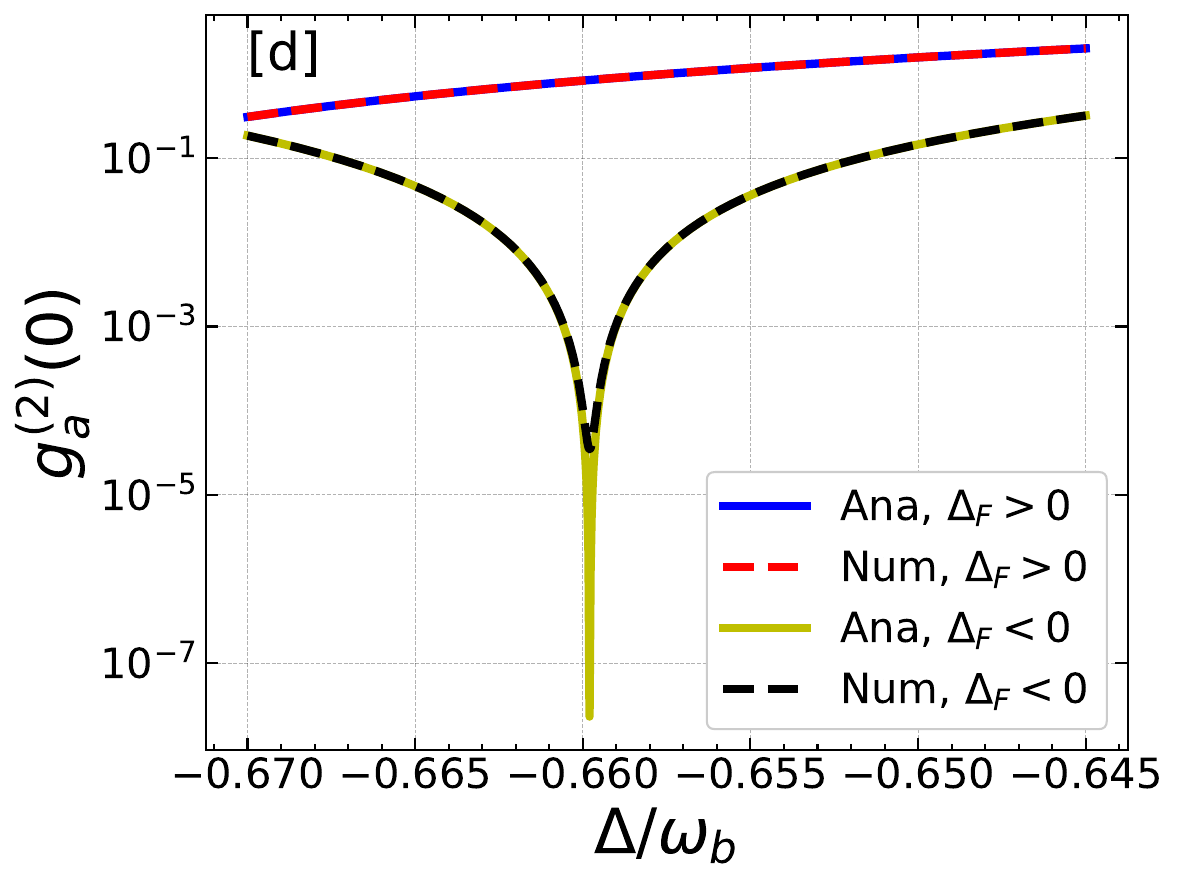}
\end{center}
\caption{Plot numerical (Num) and analytical (Ana) of the equal-time second-order correlation function $g_{\rm a}^{(2)}(0)$ versus the normalized detuning $\Delta/\omega_{\rm b}$ for $|\Delta_{\rm F}|=0.5\gamma$ with $\Lambda=\Lambda_{opt}\approx (2.46, 2.46, 2.46~~{\rm and}~~2.47)\times 10^{-6}\omega_b$ in [a], [b], [c] and [d], respectively. See the text for value of other parameters.}
\label{NThm}  
\end{figure}

Figure~\ref{NThm} compares the analytical (Ana) and numerical (Num) results of the equal-time second-order correlation function $g^{(2)}_{\rm a}(0)$ as a function of the normalized detuning $\Delta/\omega_{b}$ for $|\Delta_{F}| = 0.5\gamma$. The achieved photon blockade effect due to CW or CCW mode belong to the UCPB (destructive quantum interference). Each panel corresponds to one of the four real optimal parameter pairs $(\Delta_{\mathrm{opt}}, \Lambda_{\mathrm{opt}})$: 
(a) $\Delta_{\mathrm{opt}} = -0.684495\,\omega_{b}$, $\Lambda_{\mathrm{opt}} = 2.46157\times10^{-6}\omega_{b}$; 
(b) $\Delta_{\mathrm{opt}} = 0.654639\,\omega_{b}$, $\Lambda_{\mathrm{opt}} = 2.45563\times10^{-6}\omega_{b}$; 
(c) $\Delta_{\mathrm{opt}} = 0.679535\,\omega_{b}$, $\Lambda_{\mathrm{opt}} = 2.46105\times10^{-6}\omega_{b}$; and 
(d) $\Delta_{\mathrm{opt}} = -0.659796\,\omega_{b}$, $\Lambda_{\mathrm{opt}} = 2.47275\times10^{-6}\omega_{b}$. 
For $\Delta_{F} = 0.5\gamma$, the CW-driven cavity clearly exhibits strong photon antibunching at the optimal detunings $\Delta_{\mathrm{opt}} = -0.684495\,\omega_{b}$ and $\Delta_{\mathrm{opt}} = 0.654639\,\omega_{b}$, where $g^{(2)}_{\rm a}(0)$ reaches values as low as $10^{-6}$. In contrast, the CCW ($\Delta_{F} = -0.5\gamma$) response under the same conditions displays photon bunching ($g^{(2)}_{\rm a}(0) \gg 10^{-6}$), confirming the nonreciprocal nature of the unconventional photon blockade (UPB), as depicted in Figs. \ref{NThm}[a] and [b]. When the rotation direction is reversed ($\Delta_{F} = -0.5\gamma$), the behavior interchanges: the CCW-driven mode shows deep antibunching at $\Delta_{\mathrm{opt}} = 0.679535\,\omega_{b}$ and $\Delta_{\mathrm{opt}} = -0.659796\,\omega_{b}$, whereas the CW-driven mode ($\Delta_{F} = 0.5\gamma$) becomes bunched, as shown in Figs. \ref{NThm}[c] and [d]. The excellent agreement between the analytical and numerical results across all four cases validates the accuracy of the truncated-state analytical model. These results collectively confirm that photon statistics in the spinning magnomechanical cavity are strongly direction-dependent, arising from destructive quantum interference between distinct excitation pathways. This interference effectively suppresses multiphoton transitions, producing single-photon emission with extremely low $g^{(2)}_{\rm a}(0)$ values. Hence, the proposed spinning microwave magnomechanical system acts as a highly tunable and nonreciprocal single-photon source capable of achieving unconventional photon blockade with high spectral purity.

This is very important to mention here that the cooperative interplay between Kerr nonlinearity, optical parametric amplification, and rotation-induced detuning asymmetry leads to the emergence of nonreciprocal unconventional photon blockade (UCPB) in our system Hamiltonian. The Kerr-type magnon nonlinearity modifies the energy structure of the two-excitation manifold which properly adjust the nonlinear phase accumulated along multiphoton excitation pathways. This is essential for satisfying the destructive quantum interference condition underlying UCPB whereas the OPA introduces a direct two-photon excitation pathway that coherently couples the vacuum state to the two-photon state. 
As a result, two competing mechanism contribute to the population of the two-photon state: (i) a sequential excitation pathway mediated by single-photon transitions and magnon-photon coupling, and (ii) a direct parametric two-photon pathway induced by the OPA. With a suitable adjustment of relative amplitudes and phases of these two pathways, destructive interference suppresses the two-photon probability which lead to strong antibunching even under weak nonlinear interaction. The rotation-induced Sagnac-Fizeau shift modifies the effective cavity detuning in a direction-dependent manner thereby altering the relative phase between these excitation pathways. This makes the destructive interference condition get satisfied only for one propagation direction whereas constructive interference occurs for the opposite direction and hence we get direction-dependent photon statistics and nonreciprocal photon blockade. Moreover, the interference condition required for suppressing $\rm C_{02}$ can be satisfied only for one propagation direction whereas it is not obtained for the opposite direction of rotation which converts interference-based photon blockade into a nonreciprocal effect. In the absence of the OPA or Kerr-induced nonlinear phase shift, this specific phase-matching condition cannot be fulfilled which shows that the cooperative action of both of these nonlinear mechanisms is most important for achieving nonreciprocal unconventional photon blockade in our system Hamiltonian.

\subsection{Effect of thermal noise}

\begin{figure}[!htb]
\begin{center}
\includegraphics[width=8cm,height=6cm]{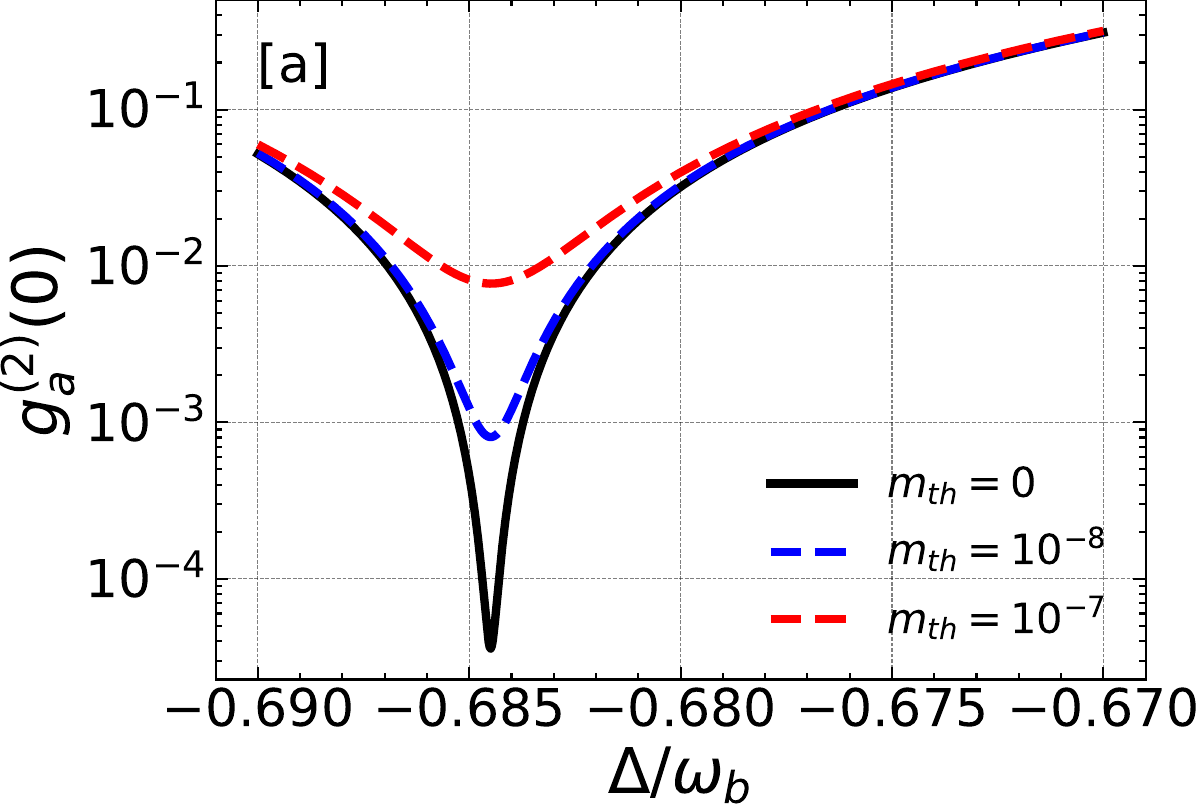}
\includegraphics[width=8cm,height=6cm]{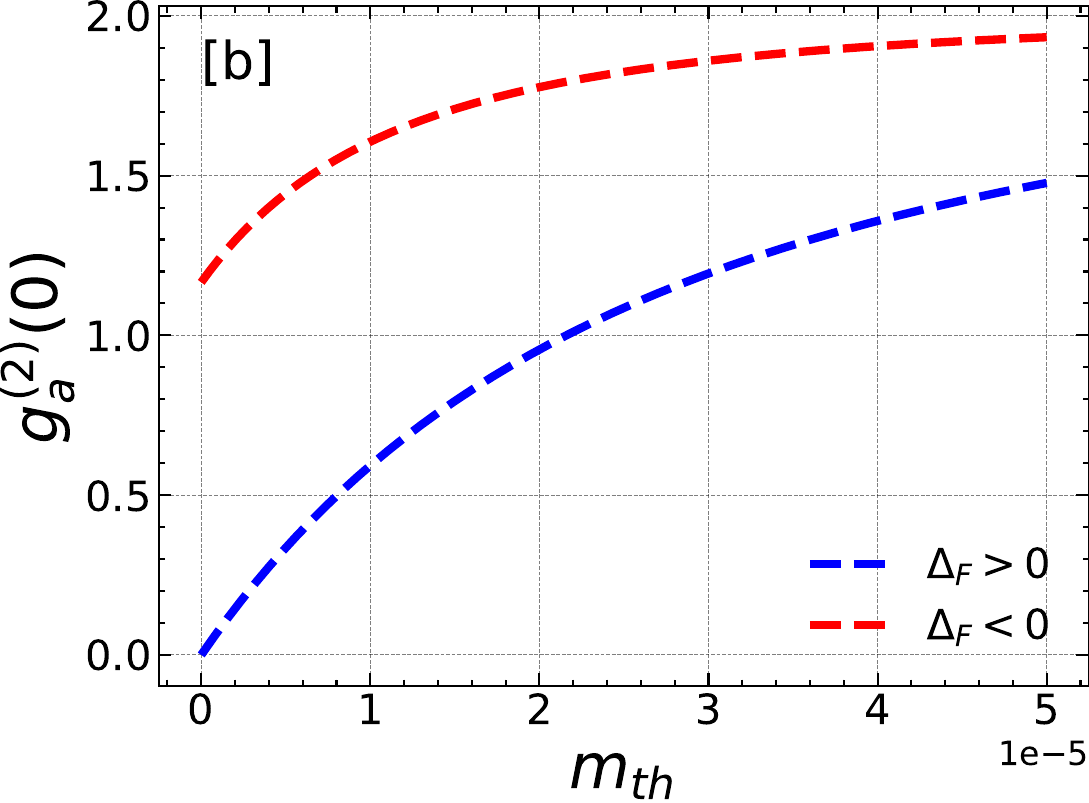}
\includegraphics[width=8cm,height=6cm]{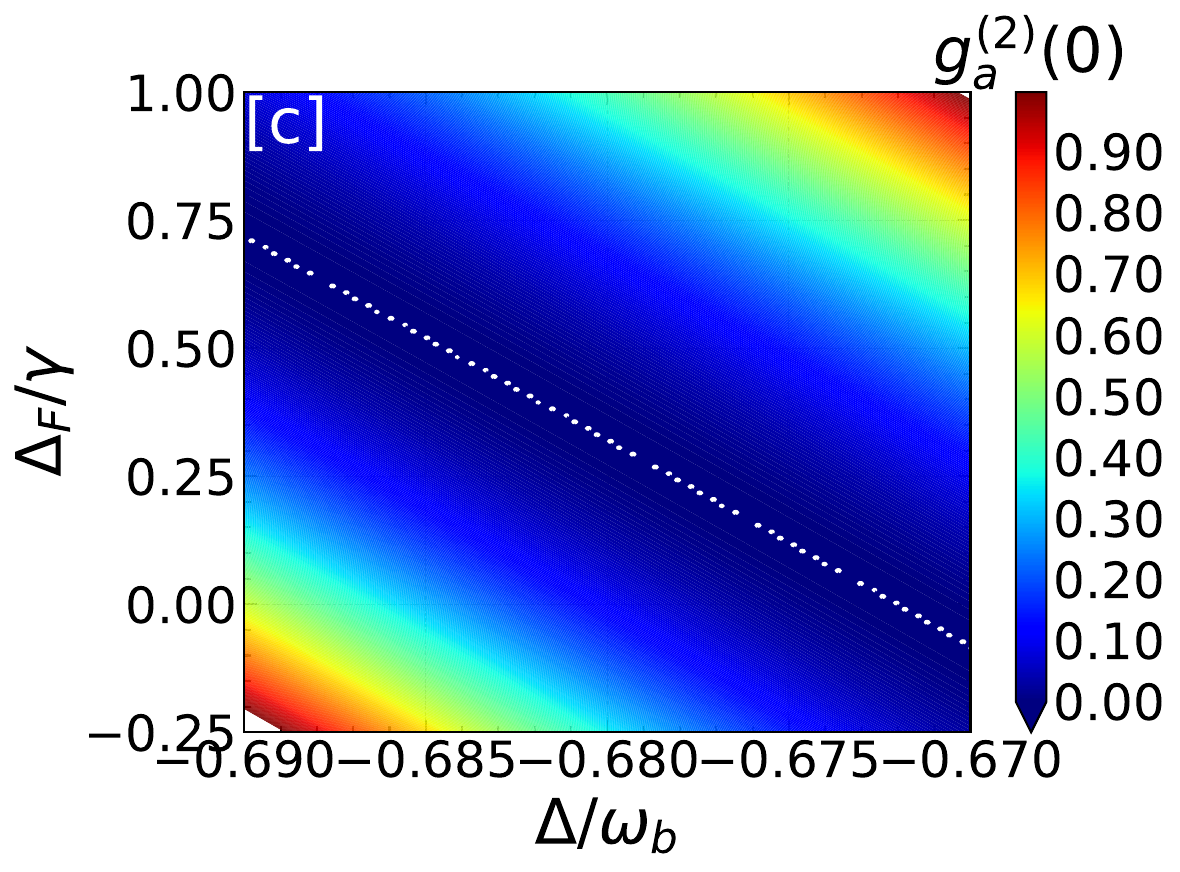}
\end{center}
\caption{Plot numerical of the equal-time second-order correlation function $g_{\rm a}^{(2)}(0)$ versus [a] the normalized detuning $\Delta/\omega_{b}$ for different values of $\rm m_{th}$ with $\Delta_F=0.5\gamma$; [b] $\rm m_{th}$ for different values of $\Delta_{\rm F}$ with $\Lambda=\Lambda_{\rm opt}\approx 2.46\times 10^{-6}\omega_b$ and $\Delta=\Delta_{\rm opt}\approx -0.68\omega_b$; and [c] the normalized $\Delta_F/\gamma$ and the normalized detuning $\Delta/\omega_{b}$, where the white dotted line indicates the optimal parameter conditions for unconventional photon blockade (UPB). See the text for value of other parameters.}
\label{NTha}  
\end{figure}

Figure~\ref{NTha} illustrates the influence of thermal noise in the magnon mode on the photon blockade behavior. In Fig.~\ref{NTha}[a], the equal-time second-order correlation function $g^{(2)}_{\rm a}(0)$ is plotted as a function of the normalized detuning $\Delta/\omega_{b}$ for different thermal magnon occupation numbers ${\rm m}_{\mathrm{th}} = 0$ (black), $10^{-8}$ (blue), and $10^{-7}$ (red). As the thermal magnon number increases, the antibunching dip in $g^{(2)}_{\rm a}(0)$ becomes progressively shallower for both CW mode, indicating that thermal excitations in the magnon bath weaken the photon blockade. Even a small decrease in ${\rm m}_{\mathrm{th}}$ significantly reduces the blockade depth, emphasizing the strong sensitivity of the system to thermal decoherence.

Figure~\ref{NTha}[b] presents $g^{(2)}_{\rm a}(0)$ as a function of the thermal magnon occupation number ${\rm m}_{\mathrm{th}}$ for both $\Delta_{F} > 0$ and $\Delta_{F} < 0$, while Fig.~\ref{NTha}[c] shows a two-dimensional density plot of $g^{(2)}_{\rm a}(0)$ versus the normalized detuning $\Delta/\omega_{b}$ and the normalized Sagnac shift $\Delta_{F}/\gamma$. In both spinning directions, $g^{(2)}_{\rm a}(0)$ increases with ${\rm m}_{\mathrm{th}}$, signifying the gradual loss of photon antibunching and the transition toward classical photon statistics. However, for $\Delta_{F} > 0$, the variation of $g^{(2)}_{\rm a}(0)$ with ${\rm m}_{\mathrm{th}}$ is relatively slow, which suggests that the CW-driven configuration is slightly more resilient to thermal noise compared with the CCW case. The density plot given in Fig.~\ref{NTha}[c] highlights that the strong antibunching occurs only near the optimal combinations of $\Delta$ and $\Delta_{F}$ (shown by a white dotted line), whereas any gradual deviations from these values rapidly suppress the photon blockade. Overall, these results demonstrate that thermal magnons play a detrimental role in realizing robust photon blockade, emphasizing the importance of maintaining low magnon temperatures for achieving stable nonreciprocal unconventional photon blockade (UCPB) in a spinning microwave magnomechanical system.

\subsection{$g_{\rm a}^{(2)}(0)$ versus $K$ and $\epsilon$}

\begin{figure}[!htb]
\begin{center}
\includegraphics[width=8cm,height=6cm]{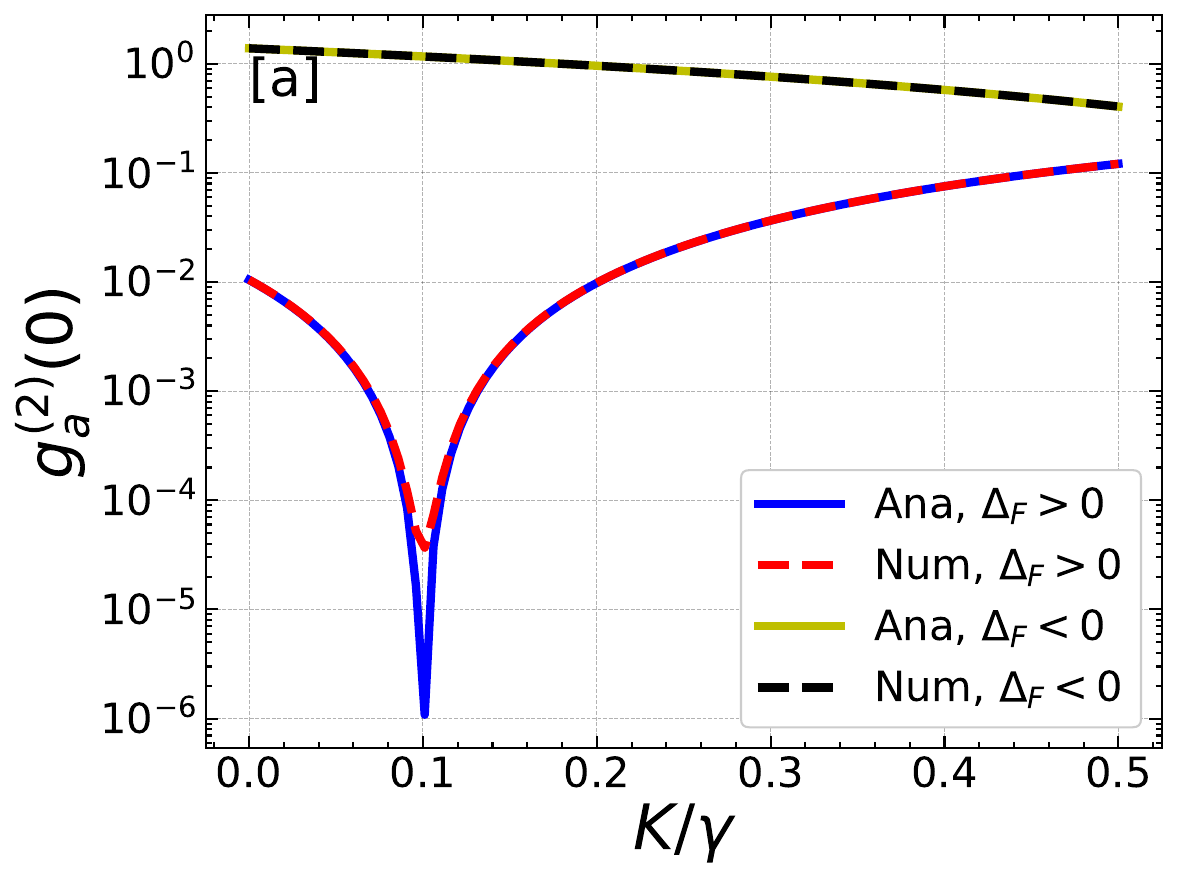}
\includegraphics[width=8cm,height=6cm]{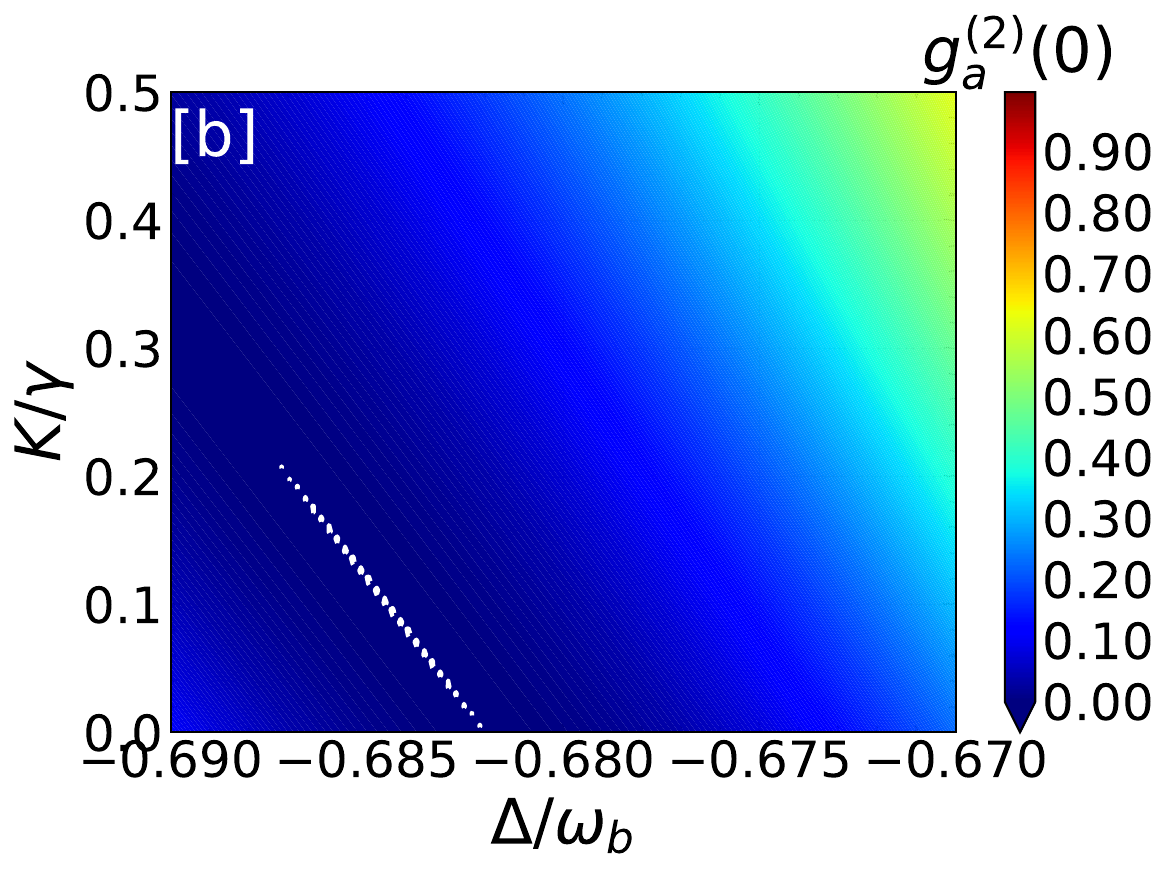}
\end{center}
\caption{Plot of the equal-time second-order correlation function $g_{\rm a}^{(2)}(0)$ parameter versus the normalized Kerr magnon parameter $K/\gamma$ with $\lambda=\Lambda_{opt}\approx 2.46\times 10^{-6}\omega_b$ and $\Delta=\Delta_{opt}=-0.68\omega_b$. Plot numerical of the $g_{\rm a}^{(2)}(0)$ as a function of the normalized detuning $\Delta/\omega_{b}$ and $K/\gamma$ with $\Lambda=\Lambda_{opt}\approx 2.46\times 10^{-6}\omega_b$, where the white dotted line indicates the optimal parameter conditions for unconventional photon blockade (UPB). See the text for value of other parameters.}
\label{fig4}  
\end{figure}

The influence of Kerr-type magnon nonlinearity on photon-blockade behavior is analyzed for both clockwise (CW) and counterclockwise (CCW) driving configurations. In Fig.~\ref{fig4}[a], the equal-time second-order correlation function $g^{(2)}_{\rm a}(0)$ is plotted as a function of the normalized Kerr parameter $K/\gamma$ for four cases: CW-driven mode Analytical (Ana) and Numerical (Num) with $\Delta_{F}>0$ (red and sky-blue curves, respectively), and CCW-driven mode Analytical (Ana) and Numerical (Num) with $\Delta_{F}<0$ (dark-blue and black curves, respectively). For the CW-driven configuration ($\Delta_{F}>0$), the analytical and numerical results overlap almost perfectly and exhibit a very strong photon antibunching minimum at $K/\gamma \approx 0.1$, where $g^{(2)}_{\rm a}(0) \sim 10^{-6}$, which confirms both the accuracy of the analytical method and the existence of robust unconventional photon blockade (UPB). In contrast, for the CCW-driven configuration ($\Delta_{F}<0$), the two curves deviate slightly and show a second pronounced antibunching minimum around $K/\gamma \approx 0.5$, also indicating strong UPB. On gradually changing  these optimal nonlinearities, $g^{(2)}_{\rm a}(0)$ increases again, reflecting the weakening of the photon blockade and the emergence of photon bunching. These results indicate that the Kerr nonlinearity facilitates the destructive interference condition for UCPB.

A density plot of $g^{(2)}_{\rm a}(0)$ as a function of both the normalized detuning $\Delta/\omega_{b}$ and the Kerr parameter $K/\gamma$ is shown in Fig.~\ref{fig4}[b] for the CCW-driven configuration ($\Delta_{F}<0$). The dark-blue regions correspond to strong photon antibunching whereas the white dotted contour marks the optimal parameter conditions for UCPB. The density plot also confirms that strong blockade occurs only within the narrow ranges of detuning and Kerr strength near these optimal points, and deviations from these regions rapidly suppress the photon blockade. These findings demonstrate that Kerr nonlinearity and rotation-induced detuning jointly determine the strength and directionality of the photon-blockade effect, enabling controllable nonreciprocal single-photon emission in the spinning microwave magnomechanical system.

\begin{figure}[!htb]
\begin{center}
\includegraphics[width=8cm,height=6cm]{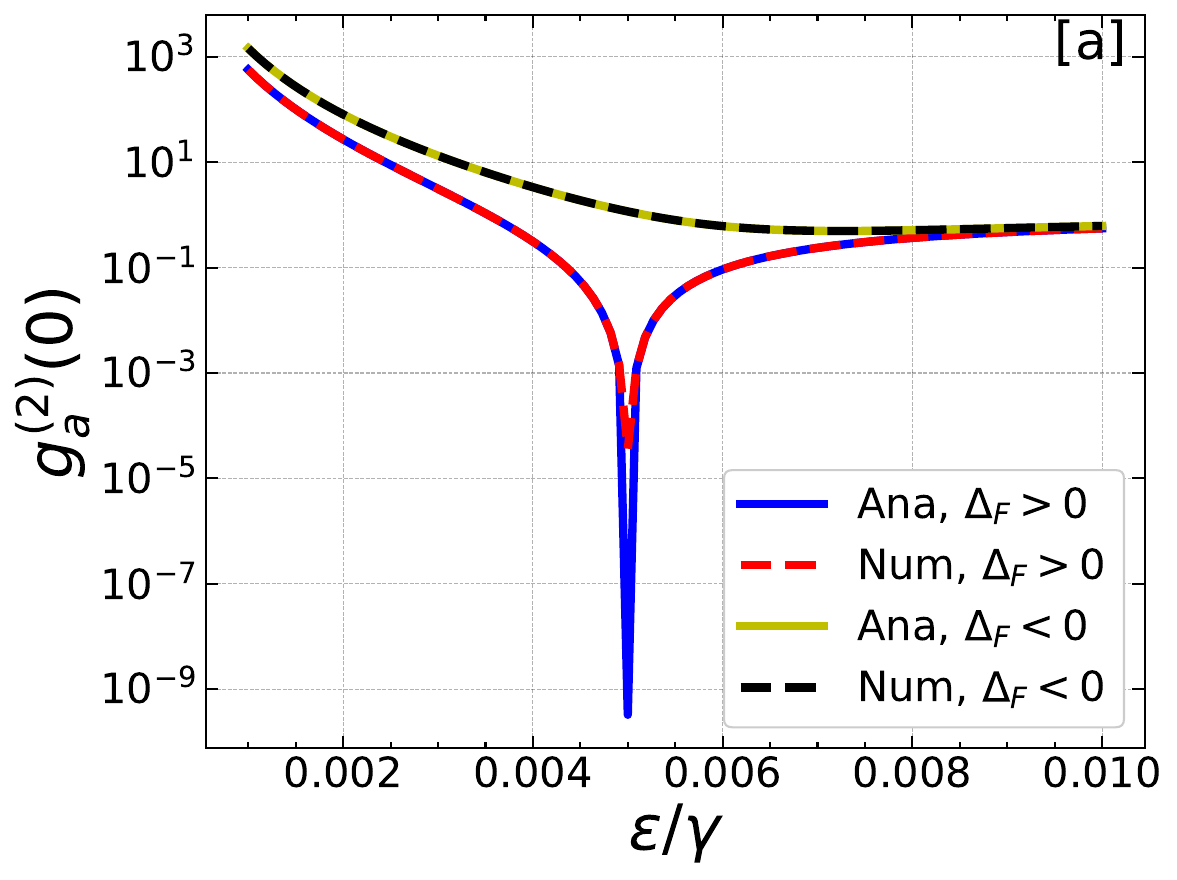}
\includegraphics[width=8cm,height=6cm]{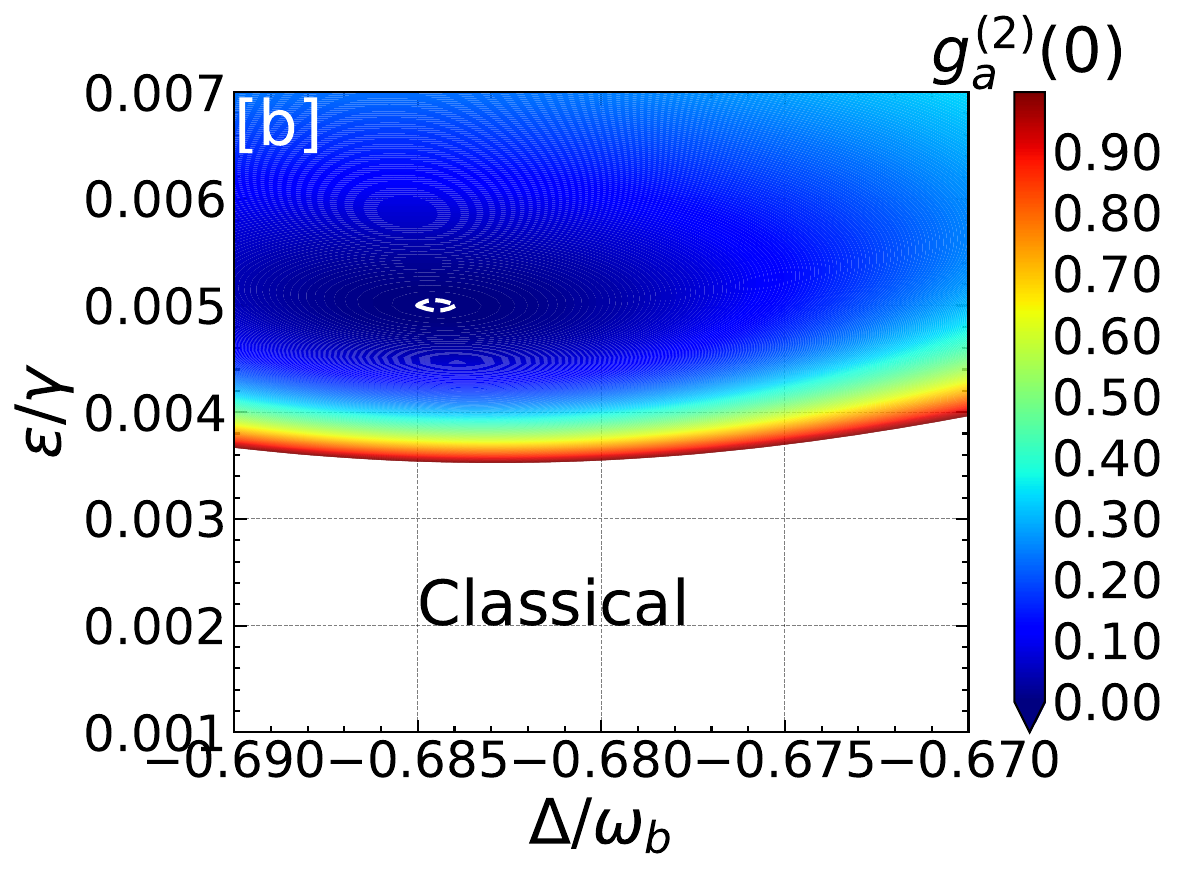}
\end{center}
\caption{Plot of the equal-time second-order correlation function $g_{\rm a}^{(2)}(0)$ parameter versus $\epsilon/\gamma$ with $\Lambda=\Lambda_{opt}\approx 2.46\times 10^{-6}\omega_b$ and $\Delta=\Delta_{opt}=-0.68\omega_b$. Plot numerical of the $g_{\rm a}^{(2)}(0)$ as a function of the normalized detuning $\Delta/\omega_{b}$ and $\epsilon/\gamma$ with $\Lambda=\Lambda_{opt}\approx 2.46\times 10^{-6}\omega_b$, where the white dotted line indicates the optimal parameter conditions for unconventional photon blockade (UPB). See the text for value of other parameters.}
\label{fig5}  
\end{figure}

The effect of the driving field amplitude on the photon-blockade behavior is analyzed for both clockwise (CW) and counterclockwise (CCW) driving configurations. In Fig.~\ref{fig5}[a], the equal-time second-order correlation function $g^{(2)}_{\rm a}(0)$ is plotted as a function of the normalized driving field amplitude $\epsilon/\gamma$ at the fixed parameters $\Lambda = \Lambda_{\mathrm{opt}} \approx 2.46\times10^{-6}\omega_{b}$ and $\Delta = \Delta_{\mathrm{opt}} = -0.68\,\omega_{b}$. For the CW-driven mode ($\Delta_{F} > 0$), the analytical and numerical results overlap closely and exhibit a very strong photon antibunching minimum at $\epsilon/\gamma \approx 0.05$, where $g^{(2)}_{\rm a}(0) \sim 10^{-6}$, confirming the realization of robust unconventional photon blockade (UPB) under weak excitation. As the driving amplitude increases beyond this range, $g^{(2)}_{\rm a}(0)$ rises sharply, indicating a transition from strong antibunching to photon bunching due to enhanced multiphoton excitations. For the CCW-driven configuration ($\Delta_{F} < 0$), a similar trend is observed, but the antibunching occurs at slightly higher $\epsilon/\gamma$ values, showing that the CW mode retains stronger blockade under weaker driving conditions. The means that OPA-induced two-photon excitation channel interferes with the sequential excitation pathway, and its strength determines the degree of cancellation of the two-photon transition amplitude responsible for unconventional photon blockade.

A two dimensional  density plot of $g^{(2)}_{\rm a}(0)$ as a function of the normalized detuning $\Delta/\omega_{b}$ and the normalised driving field amplitude $\epsilon/\gamma$ is shown in Fig.~\ref{fig5}[b]. For the CCW-driven configuration ($\Delta_{F} < 0$), it can be seen that the dark-blue regions denote strong photon antibunching whereas the white dotted contour marks the optimal parameter combinations for UCPB. It is also revealed that pronounced photon blockade occurs for $\epsilon/\gamma \geq 0.004$ and within a narrow detuning window around the optimal parameters. It can be seen that outside this range, $g^{(2)}_{\rm a}(0)$ rapidly increases, and  shifts toward white color area, which indicates  the classical photon bunching where $g^{(2)}_{\rm a}(0)\ge1$. This means that on gradually increasing the driving field amplitude beyond the threshold efficiently activates the destructive quantum interference pathways responsible for photon blockade. So, all these results highlight the crucial role of controlled driving strength for sustaining nonreciprocal single-photon emission in the spinning microwave magnomechanical system.

\subsection{Pure dephasing effects}

Since pure dephasing can undermine photon blockade and serve as a source of detrimental decoherence in the system, we investigate its specific impact on the antibunching characteristics of the cavity photons. The effects of pure dephasing are formally modeled by incorporating the following Lindblad term into the master equation:$$\mathcal{L}_p(\varrho) = \frac{\gamma_p}{2}[2{\rm a}^{\dag} {\rm a}\varrho {\rm a}^\dag {\rm a} - ({\rm a}^{\dag} {\rm a})^2 \varrho - \varrho({\rm a}^{\dag} {\rm a})^2]$$
here, $\gamma_p$ represents the pure dephasing rate associated with the cavity mode.
\begin{figure}[!htb]
\begin{center}
\includegraphics[width=8cm,height=6cm]{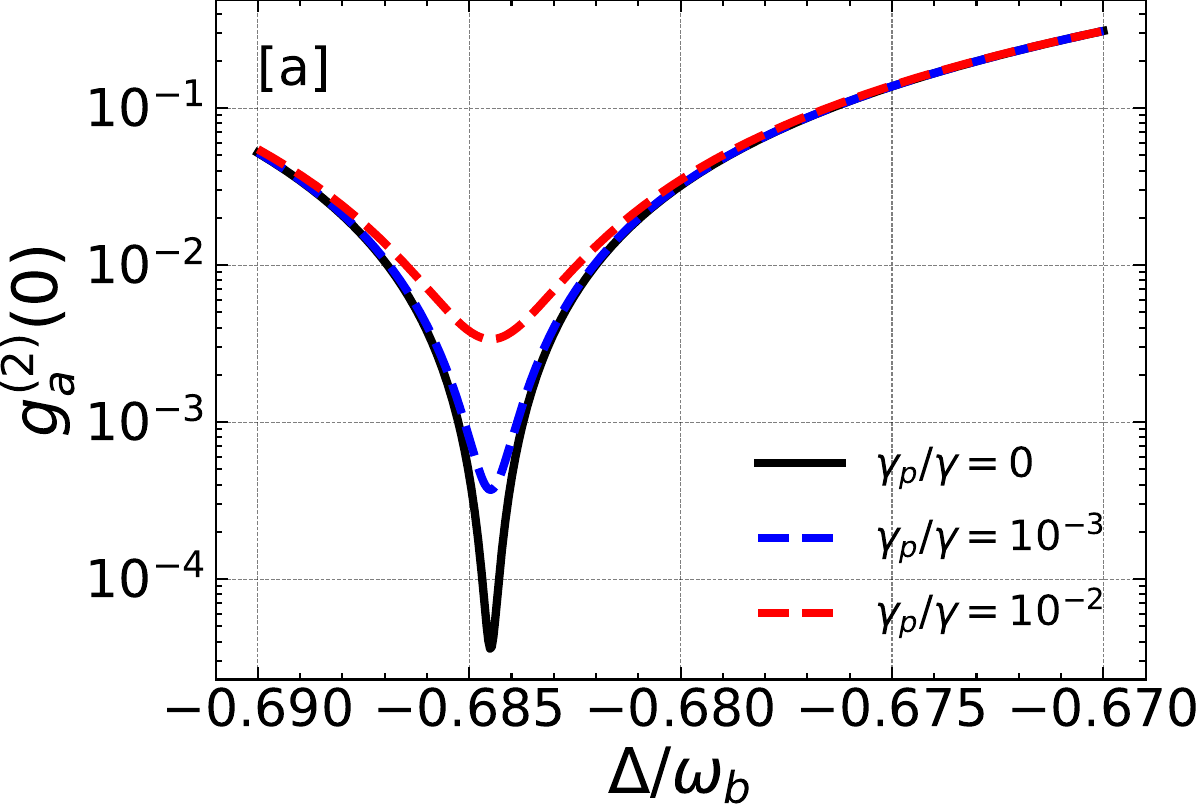}
\includegraphics[width=8cm,height=6cm]{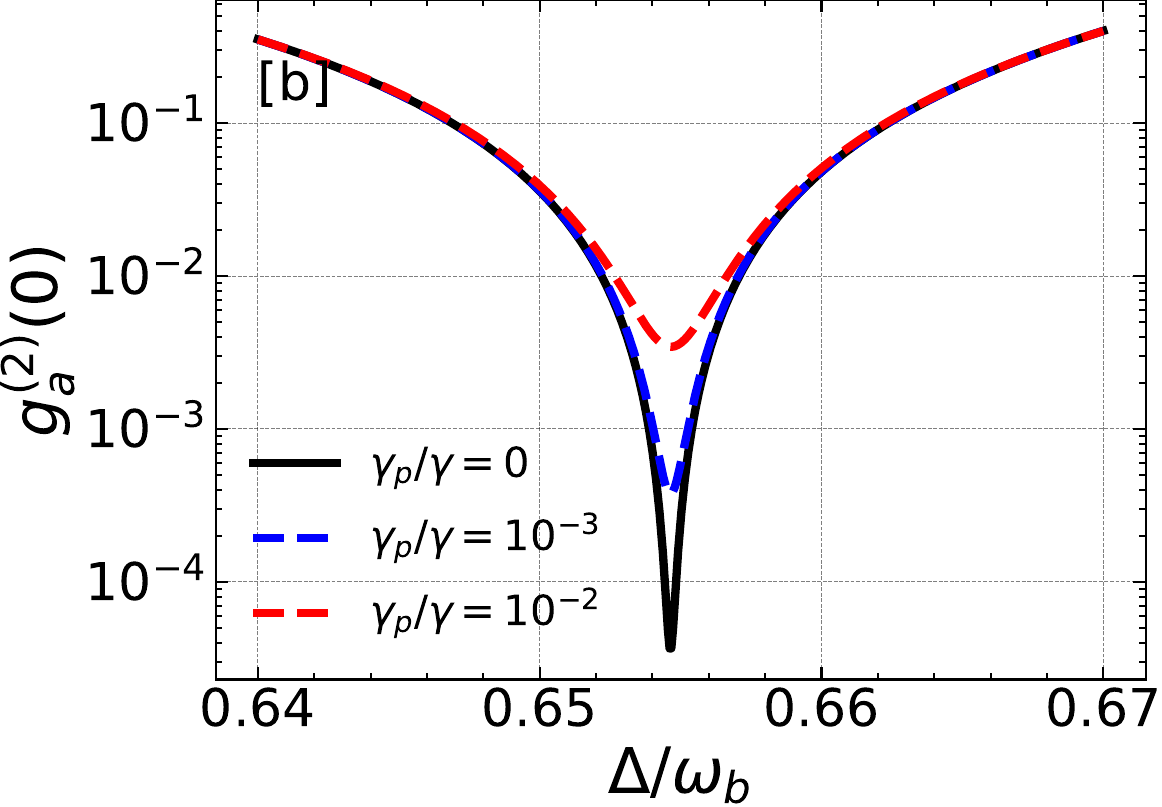}
\includegraphics[width=8cm,height=6cm]{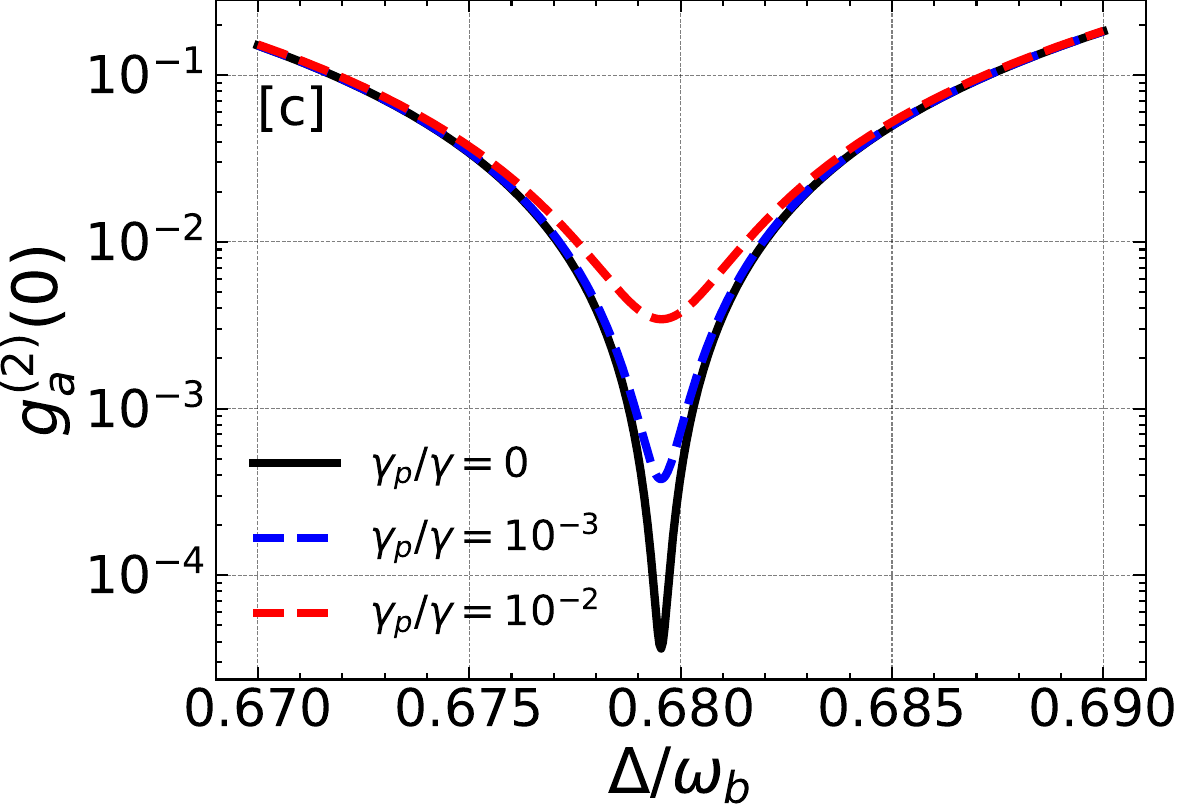}
\includegraphics[width=8cm,height=6cm]{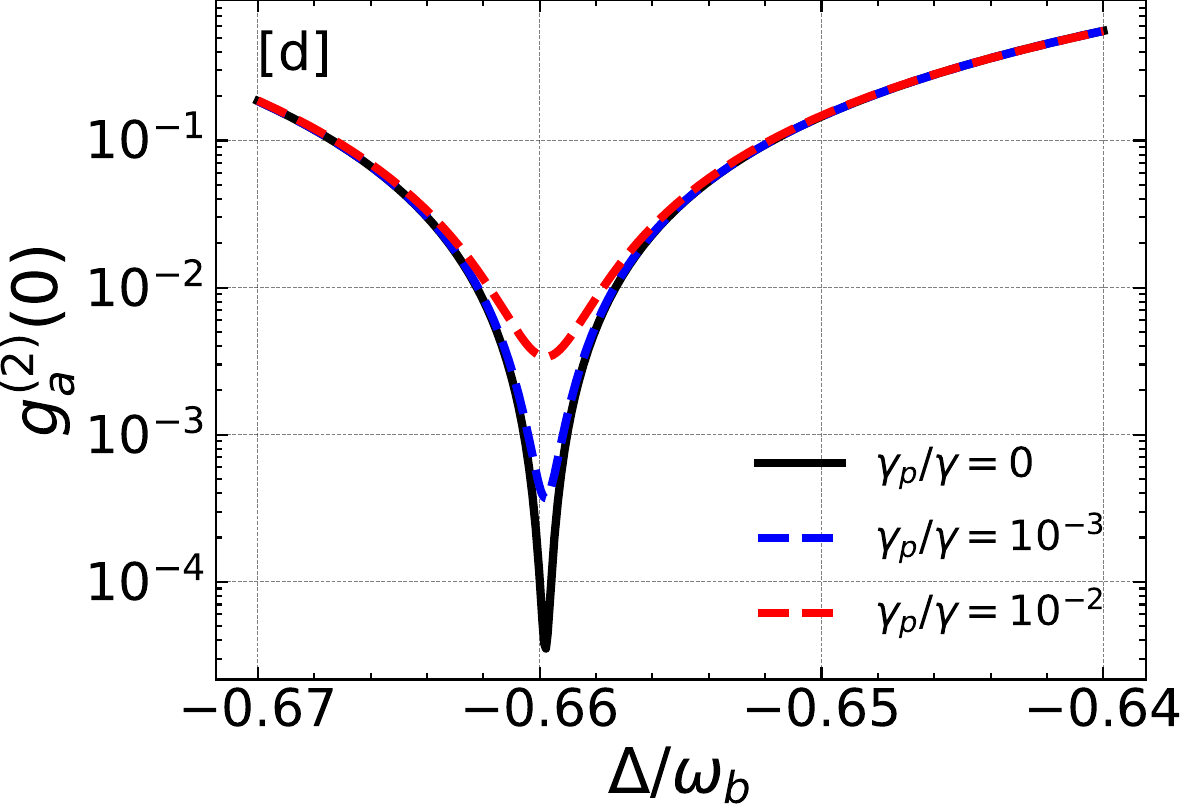}
\end{center}
\caption{Plot numerical of the equal-time second-order correlation function $g_{\rm a}^{(2)}(0)$ versus the normalized detuning $\Delta/\omega_b$ for different values of pure-dephasing rates $\gamma_p$ with $\Lambda=\Lambda_{opt}\approx (2.46, 2.46, 2.46~~{\rm and}~~2.47)\times 10^{-6}\omega_b$ in [a], [b], [c] and [d], respectively. $\Delta_{\rm F}=0.5\gamma$ in [a-b] and $\Delta_{\rm F}=-0.5\gamma$ in [c-d]. See the text for value of other parameters.}
\label{fig6}  
\end{figure}

The influence of pure dephasing on the photon-blockade characteristics is illustrated in Fig.~\ref{fig6}, which presents numerical plots of the equal-time second-order correlation function $g^{(2)}_{\rm a}(0)$ versus the normalized detuning $\Delta/\omega_{b}$ for three pure-dephasing rates $\gamma_{p}$. Panels [a]–[d] correspond to the four optimal parameter sets used previously: [a] and [b] are for $\Delta_{F}= 0.5\gamma$ (CW mode) with $\Lambda_{\mathrm{opt}}\approx2.46157\times10^{-6}\omega_{b}$ and $2.45563\times10^{-6}\omega_{b}$, respectively; [c] and [d] are for $\Delta_{F}= -0.5\gamma$ (CCW mode) with $\Lambda_{\mathrm{opt}}\approx2.46105\times10^{-6}\omega_{b}$ and $2.47275\times10^{-6}\omega_{b}$, respectively. Each panel displays multiple curves for different values of $\gamma_{p}$, permitting a direct assessment of pure-dephasing effects at the same operating point.

We get two key insights  from these plots of the equal-time second-order correlation function $g_{\rm a}^{(2)}(0)$. First, on gradually increasing pure dephasing rate uniformly weakens the photon blockade: the antibunching minima of $g^{(2)}_{\rm a}(0)$ become shallower and the values move toward unity (and above), indicating the progressive loss of nonclassical correlations. Second, the degree of robustness depends both on the chosen optimal pair $(\Delta_{\mathrm{opt}},\Lambda_{\mathrm{opt}})$ and on the driving direction. For $\Delta_{F}= 0.5\gamma$ (panels a-b), the antibunching minimum survives for small $\gamma_{p}$ although it is visibly reduced for larger dephasing rate. For the $\Delta_{F}= -0.5\gamma$ panels ([c],[d]), corresponding to the CCW-driven configuration, increasing $\gamma_{p}$ similarly weakens the photon blockade, yet the antibunching dip remains comparatively more robust for certain optimal $\Lambda_{\mathrm{opt}}$ values, persisting even at higher dephasing rates. Overall, these results demonstrate that moderate pure dephasing does not immediately destroy unconventional photon blockade (UCB)as small $\gamma_{p}$ values only reduce its depth, whereas stronger dephasing rate drives $g^{(2)}_{\rm a}(0)$ toward or above unity, marking a transition from quantum to classical photon statistics.
 
To establish a more comprehensive understanding of nonreciprocal unconventional photon blockade in our system Hamiltonian, we also explore other important complementary statistical measure beyond the equal-time correlation function. Although $g^{(2)}_{\rm a}(0)$ directly characterizes the suppression of two-photon probability, the Mandel $Q_{\rm a}$ parameter quantifies deviations from Poissonian photon statistics whereas the time-delay correlation function $g^{(2)}_{\rm a}(\tau)$ reveals the temporal dynamics of photon correlations. These indicators do not represent independent phenomena rather they offer mutually consistent tools to investigate the same rotation-induced, direction-dependent interference mechanism responsible for nonreciprocal photon antibunching. This shows that  statistical and temporal dynamics  analysis collectively reinforce the physical picture established by the equal-time correlation function.

\subsection{Non-classicality : Mandel $Q_{\rm a}$}

\begin{figure}[!htb]
\begin{center}
\includegraphics[width=8cm,height=6cm]{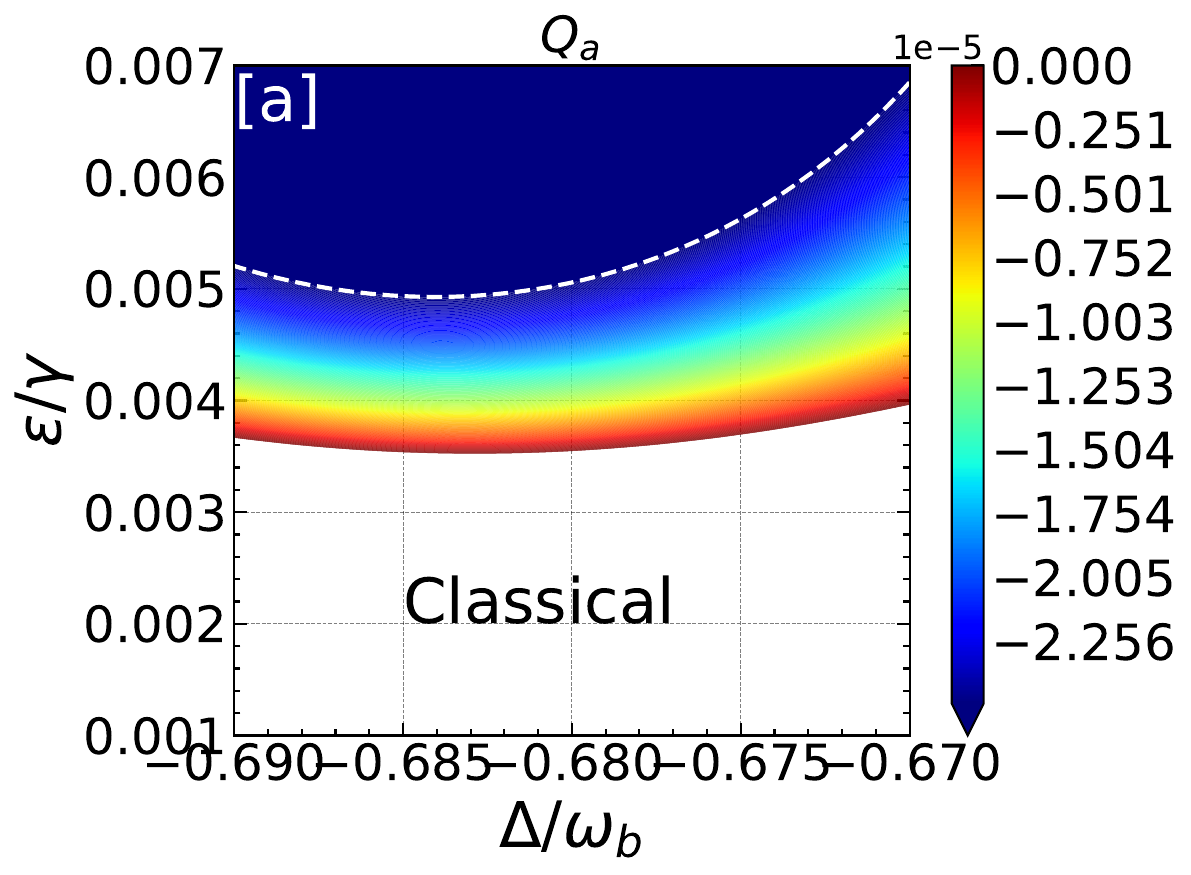}
\includegraphics[width=8cm,height=6cm]{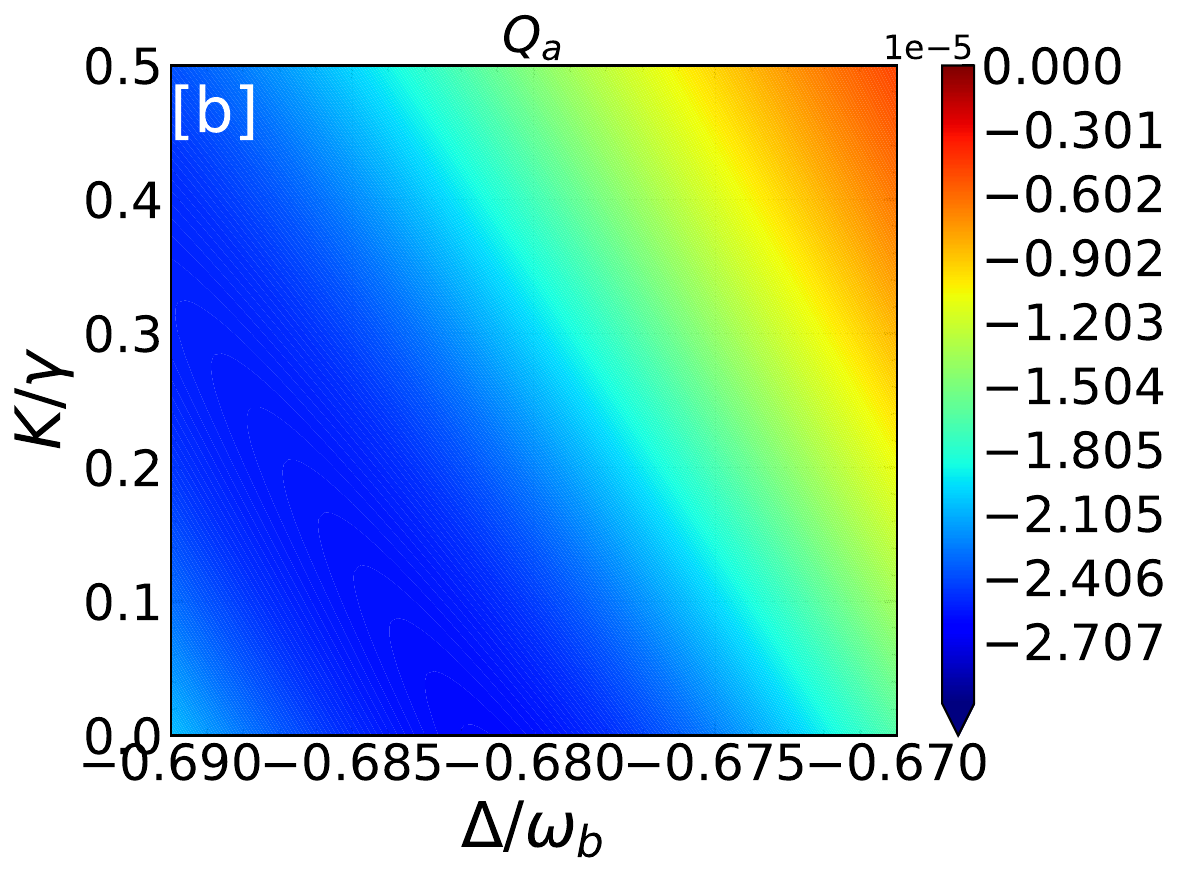}
\end{center}
\caption{Plot numerical of the Mandel $Q_{\rm a}$ parameter versus the normalized detuning $\Delta/\omega_{\rm b}$ and: $K/\gamma$ in [a] and amplitude of probe field $\epsilon/\gamma$ in [b] with $\Lambda=\Lambda_{opt}\approx 2.46\times 10^{-6}\omega_b$ and $\Delta_F=-0.5\gamma$. See the text for value of other parameters.}
\label{fig7}  
\end{figure}

To further understand the direction-dependent interference mechanism underlying the observed nonreciprocal photon blockade, we now introduce the additional statistical parameter knows as Mandel $Q_{\rm a}$ parameter as discussed in Ref.~\cite{Liang24}
\begin{equation} \label{eq:21} 
Q_{\rm a}:= \frac{\Tr(\varrho_s{\rm a^{\dag^2}}{\rm a^2})-[\Tr(\varrho_s{\rm a^{\dag}}{\rm a})]^2}{\Tr(\varrho_s{\rm a^{\dag}}{\rm a})},
\end{equation}
A negative value of $Q_{\rm a}$ ($Q_{\rm a}<0$) indicates quantum statistical properties for the photon mode, whereas a non-negative value ($Q_{\rm a}\geq0$) indicates classical properties.

The Mandel parameter $Q_{\rm a}$ quantifies deviations from Poissonian photon statistics which means the negative values ($Q_{\rm a}<0$) correspond to sub-Poissonian, nonclassical light exhibiting photon antibunching, whereas $Q_{\rm a}\ge0$ denotes Poissonian or super-Poissonian (classical or bunched) statistics. Both subplots of Figure~\ref{fig7} correspond to the CCW-driven configuration ($\Delta_{F}=-0.5\gamma$) with the optimal coupling $\Lambda_{\mathrm{opt}}\approx2.46\times10^{-6}\omega_{b}$.

Figure~\ref{fig7}[a] shows $Q_{\rm a}$ as a function of the normalized detuning $\Delta/\omega_{b}$ and the driving field amplitude $\epsilon/\gamma$. A pronounced dark-blue region with $Q_{\rm a}<0$ appears for $\epsilon/\gamma \geq 0.005$ near the optimal detuning, revealing strong sub-Poissonian photon statistics consistent with unconventional photon blockade (UCPB). The white dotted line indicates the optimal parameter combinations where destructive quantum interference maximizes photon blockade whereas outside this narrow region, $Q_{\rm a}$ approaches  towards zero, signifying a transition to classical photon bunching.

Figure~\ref{fig7}[b] presents $Q_{\rm a}$ as a function of the normalized detuning $\Delta/\omega_{b}$ and the Kerr nonlinearity $K/\gamma$. It can be seen that the negative-$Q_{\rm a}$ regions are observed around $K/\gamma\approx0.5$ and near the same optimal detuning, confirming that photon blockade generation is highly sensitive to both Kerr strength and detuning. The white or light-coloured regions in both subplots correspond to $Q_{\rm a}$ approaching towards zero which indicate classical photon statistics. Overall, the Mandel parameter behavior very closely parallels that of $g^{(2)}_{\rm a}(0)$ which means that regions with $Q_{\rm a}<0$ coincide with $g^{(2)}_{\rm a}(0)<1$, confirms that both consistently identify the unconventional photon-blockade regime and its dependence on driving amplitude, Kerr nonlinearity strengths, and cavity detuning in the spinning microwave magnomechanical system. This direction-dependent sub-Poissonian behavior again confirms that the suppression of multiphoton probability originates from the same interference mechanism which is responsible for nonreciprocal photon blockade.

\subsection{Time evolution of the second-order correlation function: $g_{\rm a}^{(2)}(\tau)$}

\begin{figure}[!htb]
\begin{center}
\includegraphics[width=8cm,height=6cm]{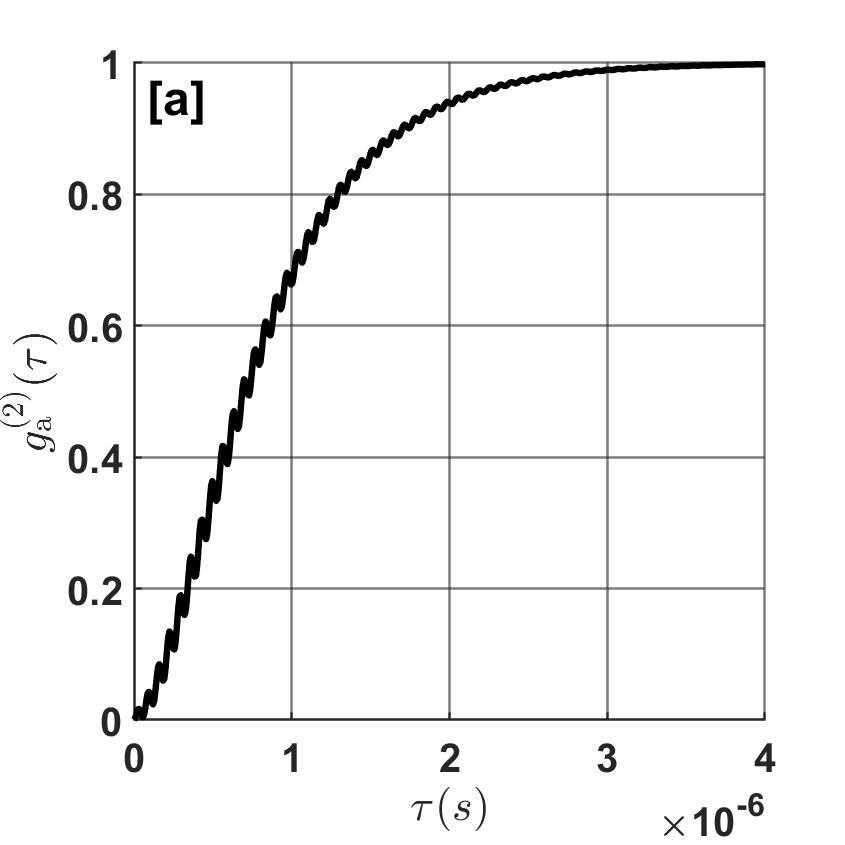}
\includegraphics[width=8cm,height=6cm]{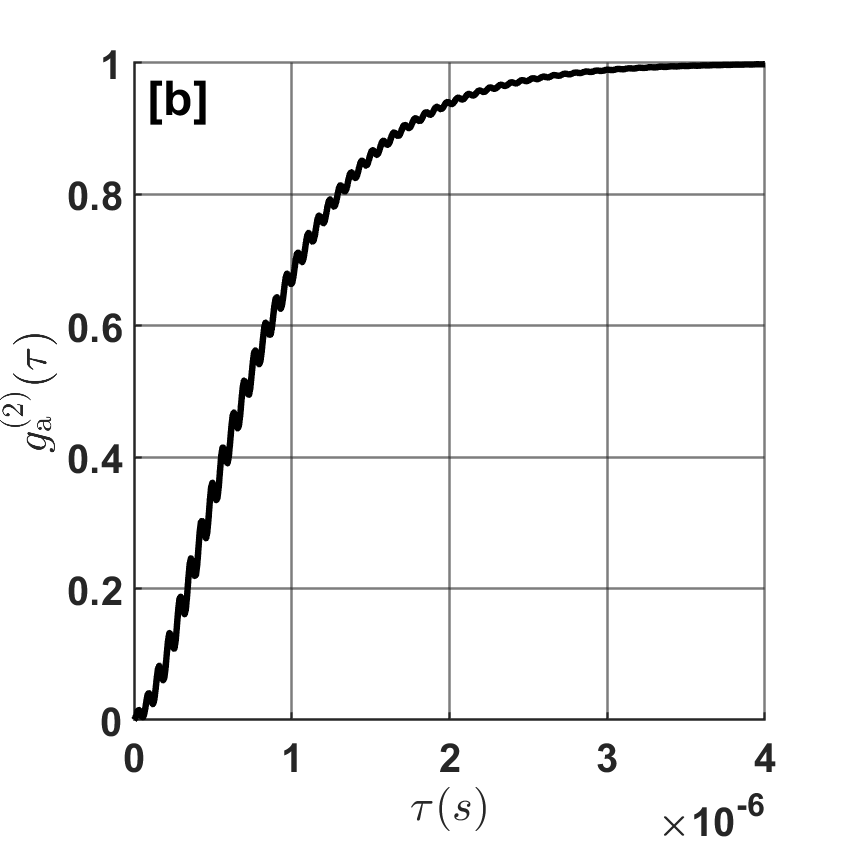}
\includegraphics[width=8cm,height=6cm]{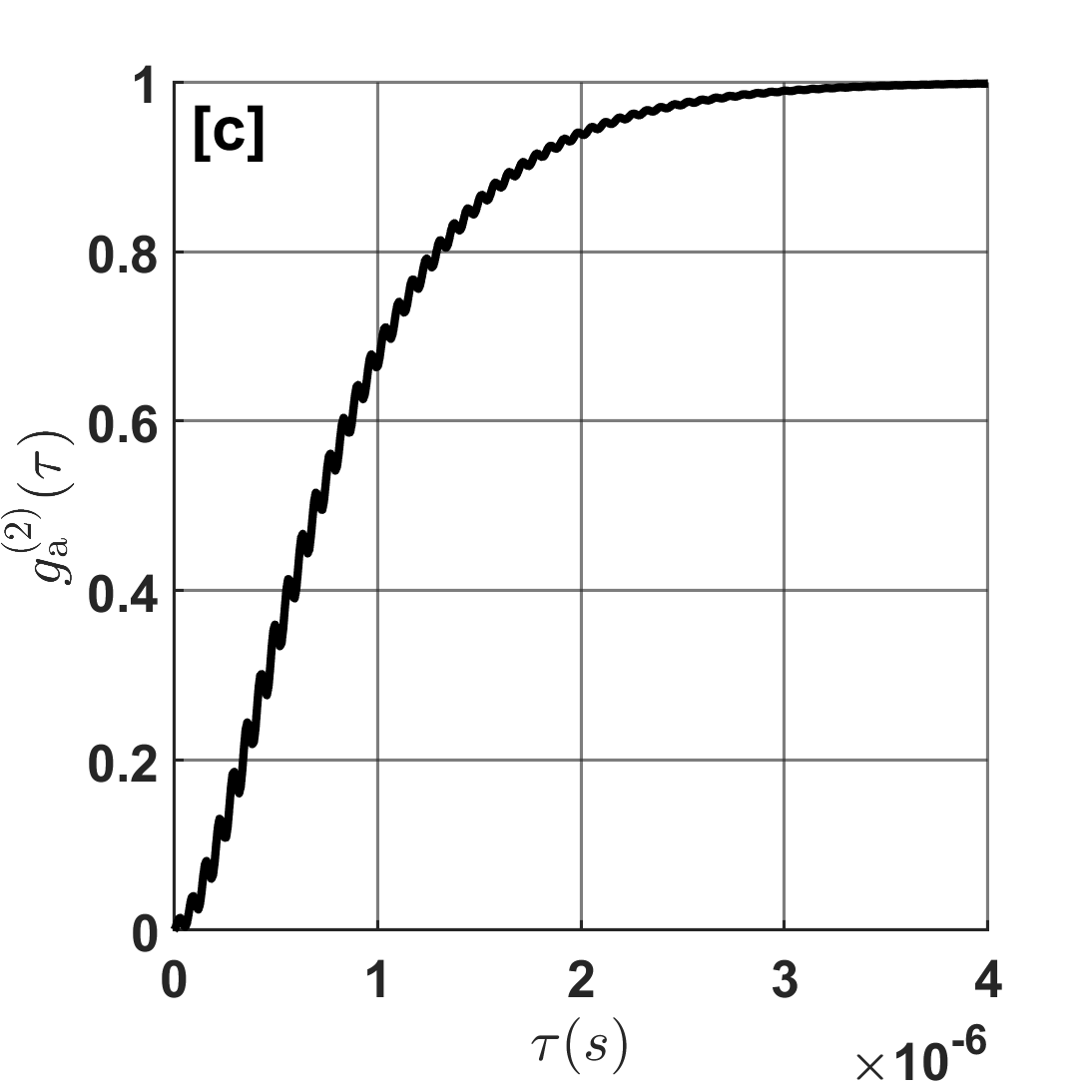}
\includegraphics[width=8cm,height=6cm]{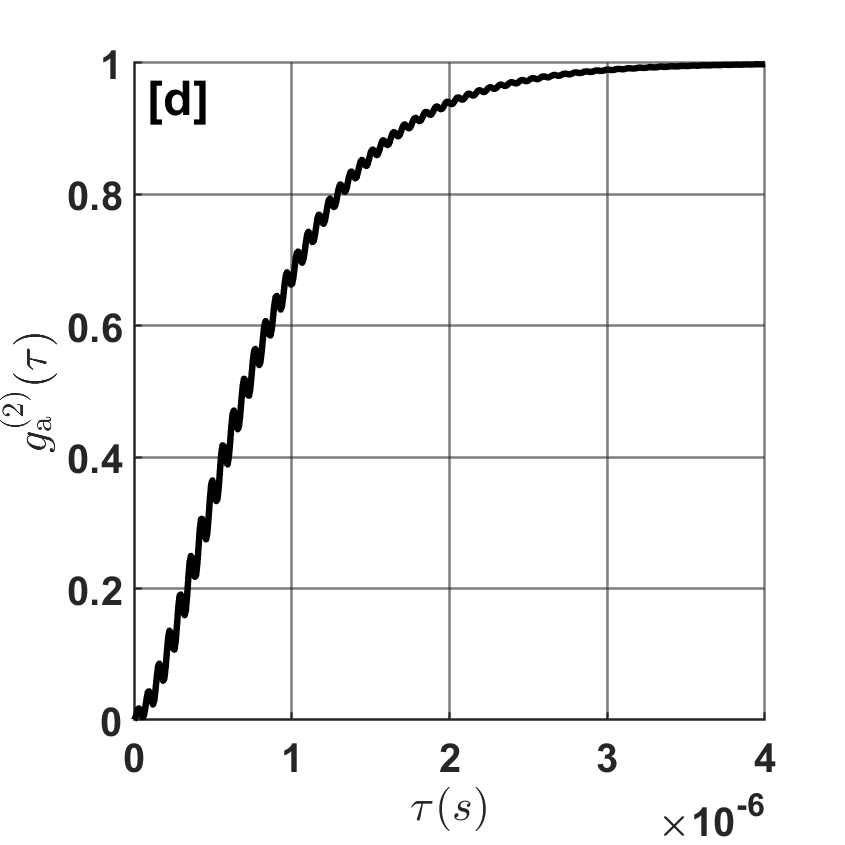}
\end{center}
\caption{Time evolution of the second-order correlation function, $g_{\rm a}^{(2)}(\tau)$ $(t = 0)$ with $\Delta_F=+0.5\gamma$ in [a] and [b] with $\Lambda= 2.45563\times 10^{-6}\omega_b$ and $\Lambda= 2.46157\times 10^{-6}\omega_b$, respectively; and $\Delta_F=-0.5\gamma$ in [c] and [d] with $\Lambda= 2.46105\times 10^{-6}\omega_b$ and $\Lambda= 2.47275\times 10^{-6}\omega_b$, respectively. See the text for value of other parameters.}
\label{fig8}  
\end{figure}

The second-order correlation function, $g_{\rm a}^{(2)}(\tau)$, is a significant quantity for characterizing the statistical properties of the photons. It represents the joint probability of detecting a photon at time $t$ and a second photon at time $t+\tau$ (up to a proportionality constant), and its time evolution is calculated as
\begin{equation} \label{eq:21}
g_{\rm a}^{(2)}(\tau)=\frac{\langle {\rm a}^\dag (t) {\rm a}^\dag(t+\tau){\rm a}(t+\tau){\rm a} (t) \rangle}{\langle {\rm a}^\dag (t){\rm a} (t) \rangle^2}
\end{equation}

The condition $g_{\rm a}^{(2)}(\tau) > g_{\rm a}^{(2)}(0)$ for delay times $\tau > 0$ provides evidence of photon blockade and confirms the sub-Poissonian statistics of the emitted photons. We note that $g_{\rm a}^{(2)}(\tau)$ grows with increasing $\tau$, eventually reaching its maximum value (unity) near $\tau \approx 3.5\,\mu\text{s}$, as illustrated in Figure \ref{fig8} [a-d]. Furthermore, $g_{\rm a}^{(2)}(\tau)$ increases until it reaches its maximum value of unity (characteristic of a coherent state) at $\tau \approx 3.5\,\mu\text{s}$ (the approximate lifetime of the photons in the cavity), as shown in Figure \ref{fig8} [a-d]. This indicates that as the time delay increases, the photons eventually become statistically independent (coherent). 

\section{Experimental feasibility}

The proposed scheme is shown to be experimentally feasible, consistent with recent experimental benchmarks. Following \cite{Xu24}, we set the magnon mode frequency to $\omega_{m}/2\pi = 10.1$ GHz. In accordance with \cite{Xu24}, we adopt a magnon dissipation rate of $\gamma_{\rm m}/2\pi = 0.55$ MHz. Furthermore, the generation of magnon squeezing has been experimentally verified \cite{Shen22}, offering a robust resource for precision measurements. The experimental parameters for the spinning resonator are selected as ${\rm r} = 30\,\mu\text{m}$ \cite{Hou25}, ${\rm n} = 1.4$ \cite{Hou25}, $\omega_{a}/2\pi = 10.1\,\text{GHz}$ \cite{LiuOL19}, and $\gamma_{\rm m}/2\pi = 1.1\,\text{MHz}$ \cite{LiuIEEE19}, with a corresponding rotation speed of $\Omega \approx 6.43 \times 10^{2}\,\text{MHz}$.

\section{Conclusion}

In summary, we realize nonreciprocal unconventional photon blockade in a spinning microwave cavity magnomechanical system. This is made possible by combining the nonlinear effects from Kerr-magnon interactions and a degenerate optical parametric amplifier under weak pump driving. The rotation-induced Sagnac–Fizeau shift plays a key role in enabling the directional photon blockade, as confirmed through both analytical theory and numerical simulation. Our results show that the effect remains highly robust against thermal noise, and we examine how the probe field amplitude and magnetic-dipole coupling strength influence the system response. We also demonstrate the nonclassical behavior of the magnon and photon modes using the Mandel parameter, along with the time-delay second-order correlation function. We clarify that Kerr nonlinearity and the optical parametric amplifier play complementary roles in generating this effect. The Kerr interaction primarily controls the nonlinear energy shift and phase matching condition required for destructive interference, whereas the OPA provides an additional coherent two-photon excitation pathway necessary for interference-based photon blockade. The rotation-induced Sagnac-Fizeau shift then converts this interference mechanism into a direction-dependent effect by asymmetrically modifying the effective detuning for clockwise and counter-clockwise modes. Compared with existing nonreciprocal photon blockade platforms that typically rely on strong intrinsic nonlinearity or engineered dissipation asymmetry, our scheme achieves direction-dependent antibunching through controllable rotation induced phase asymmetry combined with interference based photon blockade. Overall, this work offers a promising pathway toward implementing direction-dependent single-photon sources and advancing integrated quantum technologies in spinning cavity-magnonic systems. 



\section*{Acknowledgments}

The research work was supported by Princess Nourah bint Abdulrahman University Researchers Supporting Project number (PNURSP2026R157), Princess Nourah bint Abdulrahman University, Riyadh, Saudi Arabia. The authors are thankful to the Deanship of Graduate Studies and Scientific Research at University of Bisha for supporting this work through the Fast-Track Research Support Program.


\begin{thebibliography}{2}

\bibitem{Giovannetti2011} V. Giovannetti, S. Lloyd, and L. Maccone, Nat. Photonics \textbf{5}, 222 (2011).
\bibitem{Stannigel2012} K. Stannigel \textit{et al.}, Phys. Rev. Lett. \textbf{109}, 013603 (2012).
\bibitem{Bennett2000} C. H. Bennett and D. P. DiVincenzo, Nature \textbf{404}, 247 (2000).
\bibitem{Buluta2011} I. Buluta, S. Ashhab, and F. Nori, Rep. Prog. Phys. \textbf{74}, 104401 (2011).
\bibitem{Faraon2008} A. Faraon \textit{et al.}, Nat. Phys. \textbf{4}, 859 (2008).
\bibitem{Birnbaum2005} K. M. Birnbaum \textit{et al.}, Nature \textbf{436}, 87 (2005).
\bibitem{Reinhard2012} A. Reinhard \textit{et al.}, Nat. Photonics \textbf{6}, 93 (2012).
\bibitem{Muller2015} K. M\"{u}ller \textit{et al.}, Phys. Rev. Lett. \textbf{114}, 233601 (2015).
\bibitem{Hamsen2017} C. Hamsen \textit{et al.}, Phys. Rev. Lett. \textbf{118}, 133604 (2017).
\bibitem{Zheng2023} C. M. Zheng \textit{et al.}, New J. Phys. \textbf{25}, 043030 (2023).
\bibitem{Lang2011} C. Lang \textit{et al.}, Phys. Rev. Lett. \textbf{106}, 243601 (2011).
\bibitem{Hoffman2011} A. J. Hoffman \textit{et al.}, Phys. Rev. Lett. \textbf{107}, 053602 (2011).
\bibitem{Vaneph2018} C. Vaneph \textit{et al.}, Phys. Rev. Lett. \textbf{121}, 043602 (2018).
\bibitem{Snijders2018} H. J. Snijders \textit{et al.}, Phys. Rev. Lett. \textbf{121}, 043601 (2018).
\bibitem{Sayrin2015} C. Sayrin \textit{et al.}, Phys. Rev. X \textbf{5}, 041036 (2015).
\bibitem{Tang2019} L. Tang \textit{et al.}, Phys. Rev. A \textbf{99}, 043833 (2019).
\bibitem{Kamal2011} A. Kamal, J. Clarke, and M. H. Devoret, Nat. Phys. \textbf{7}, 311 (2011).
\bibitem{Sounas2017} D. L. Sounas and A. Alù, Nat. Photonics \textbf{11}, 774 (2017).
\bibitem{Svela2020} A. Ø. Svela \textit{et al.}, Light Sci. Appl. \textbf{9}, 204 (2020).
\bibitem{Jiao2020} Y. F. Jiao \textit{et al.}, Phys. Rev. Lett. \textbf{125}, 143605 (2020).
\bibitem{Jiao2022} Y. F. Jiao \textit{et al.}, Phys. Rev. Appl. \textbf{18}, 064008 (2022).
\bibitem{Ren2022} Y. L. Ren, Opt. Lett. \textbf{47}, 1125 (2022).
\bibitem{Chen2023} J. Chen \textit{et al.}, Phys. Rev. B \textbf{108}, 024105 (2023).
\bibitem{Jiang2018} Y. Jiang \textit{et al.}, Phys. Rev. Appl. \textbf{10}, 064037 (2018).
\bibitem{Xu2021} Y. Xu \textit{et al.}, Phys. Rev. A \textbf{103}, 053501 (2021).
\bibitem{Tabuchi14} Y. Tabuchi, S. Ishino, T. Ishikawa, R. Yamazaki,
K. Usami, and Y. Nakamura. Phys. Rev. Lett. 113, 083603 (2014).
\bibitem{Li18} Jie Li, S-Y Zhu and G. S. Agarwal. Phys. Rev. Lett. 121, (2018) 203601.
\bibitem{Li20} Y. Mei, H. Shen and Jie Li. Phys. Rev. Lett. 124, (2020) 213604.
\bibitem{Zhang16} X. Zhang, C. L. Zou, L. Jiang, and H. X. Tang. Sci. Adv. 2, e1501286 (2016).
\bibitem{Tabuchi15} Y. Tabuchi, S. Ishino, A. Noguchi, T. Ishikawa, R. Yamazaki, K. Usami, and Y. Nakamura. Science, 349(6246), 405-408 (2015).
\bibitem{Lachance20} D. Lachance-Quirion, S. P. Wolski, Y. Tabuchi, S. Kono, K. Usami, and Y. Nakamura. Science 367.6476 (2020) 425-428.
\bibitem{FB23} M. Amazioug, B. Teklu and M. Asjad. Scientific Reports 13.1 (2023) 3833.
\bibitem{FBEntropy} M. Amazioug, S. K. Singh, B. Teklu and M. Asjad. Entropy 25.10 (2023) 1462.
\bibitem{FB-EPJP} M. Amazioug, J.-X. Peng, D. Dutykh, M. Asjad, Eur. Phys. J. Plus 140 (2025) 132.
\bibitem{SR25} F. H. Mathkoor, S. K. Singh, R. Ahmed, J. X. Peng, M. Amazioug, M. Khalid and A. Sohail. Scientific Reports, 15(1), 13503 (2025).
\bibitem{Asjad23} M. Asjad, J. Li, S. Y. Zhu and J. Q. You. Fundamental Research, 3(1) (2023) 3-7.
\bibitem{mPB} M. Amazioug, D. Dutykh, B. Teklu and M. Asjad. Annalen der Physik 536.4 (2024) 2300357.
\bibitem{Xu24} X. Deng, K. K. Zhang, T. Shui and W. X. Yang. Physical Review A, 110(6) (2024) 063711.
\bibitem{Hou25} R. Hou, W. Zhang, X. Han, H. F. Wang and S. Zhang. Scientific Reports, 15(1) (2025) 5145.
\bibitem{Ge25} P. C. Ge, Y. Yu, H. T. Wu, X. Han, H. F. Wang, and S. Zhang. Scientific Reports, 15(1) (2025) 7937.
\bibitem{Zhang25} W. Zhang, S. Liu, S. Zhang, H. F. Wang. Optics Express, 33(2) (2025) 3339-3349.
\bibitem{Huang18} R. Huang, A. Miranowicz, J.-Q. Liao, F. Nori, and H. Jing, Phys. Rev. Lett. 121, 153601 (2018).
\bibitem{Yang25} J. X. Yang, Cheng Shang, Yan-Hui Zhou, and H. Z. Shen, Simultaneous nonreciprocal unconventional photon blockade via two degenerate optical parametric amplifiers in spinning resonators, arXiv: 2505.10255 (2025).
\bibitem{DFDB25} M. Amazioug, S. Abdel-Khalek and M. Asjad. Annalen der Physik. 537(12), e00289 (2025).
\bibitem{Amghar24} M. Amghar and M. Amazioug. Optik 311 (2024) 171940
\bibitem{Edet24} C. O. Edet, M. Asjad, D. Dutykh, N. Ali and O. Abah. Physical Review Research, 6(3) (2024) 033037.
\bibitem{Deng24} X. Deng, K. K. Zhang, T. Shui, W. X. Yang, Phys. Rev. A 2024, 110, 063711.
\bibitem{Ebrahimi23} M. S. Ebrahimi and M. B. Harouni. Journal of Physics B: Atomic, Molecular and Optical Physics, 56(23), 235501 (2023).

\bibitem{Mirza2019} I. M. Mirza \textit{et al.}, Opt. Express \textbf{27}, 25515 (2019).
\bibitem{Peng2023} M. Peng \textit{et al.}, Phys. Rev. A \textbf{107}, 033507 (2023).
\bibitem{Li2021} B. Li \textit{et al.}, Phys. Rev. A \textbf{103}, 053522 (2021).
\bibitem{Huang2018} R. Huang \textit{et al.}, Phys. Rev. Lett. \textbf{121}, 153601 (2018).
\bibitem{Wang2019} K. Wang \textit{et al.}, Phys. Rev. A \textbf{100}, 053832 (2019).
\bibitem{Shen2020} H. Z. Shen \textit{et al.}, Phys. Rev. A \textbf{101}, 013826 (2020).
\bibitem{Xu2020} X. W. Xu \textit{et al.}, Phys. Rev. Appl. \textbf{13}, 044070 (2020).
\bibitem{Shang2021} X. Shang \textit{et al.}, Laser Phys. Lett. \textbf{18}, 115202 (2021).
\bibitem{Liu2023} Y. M. Liu \textit{et al.}, Opt. Express \textbf{31}, 12847 (2023).
\bibitem{Li2019} B. Li \textit{et al.}, Photonics Res. \textbf{7}, 630 (2019).
\bibitem{Liu2023a} Y. M. Liu \textit{et al.}, Phys. Rev. A \textbf{107}, 063701 (2023).
\bibitem{Zhang2023} W. Zhang \textit{et al.}, Sci. China Phys. Mech. Astron. \textbf{66}, 240313 (2023).
\bibitem{Xue2020} W. S. Xue \textit{et al.}, Opt. Lett. \textbf{45}, 4424 (2020).
\bibitem{Jing2021} Y. W. Jing \textit{et al.}, Phys. Rev. A \textbf{104}, 033707 (2021).
\bibitem{Xie2022} H. Xie \textit{et al.}, Phys. Rev. A \textbf{106}, 053707 (2022).
\bibitem{Kittel1948} C. Kittel, Phys. Rev. \textbf{73}, 155 (1948).
\bibitem{Zhang2014} X. Zhang \textit{et al.}, Phys. Rev. Lett. \textbf{113}, 156401 (2014).
\bibitem{Huebl2013} H. Huebl \textit{et al.}, Phys. Rev. Lett. \textbf{111}, 127003 (2013).
\bibitem{Bai2015} L. Bai \textit{et al.}, Phys. Rev. Lett. \textbf{114}, 227201 (2015).
\bibitem{Kippenberg2005} T. J. Kippenberg \textit{et al.}, Phys. Rev. Lett. \textbf{95}, 033901 (2005).
\bibitem{Zhao2020} W. Zhao \textit{et al.}, Sci. China Phys. Mech. Astron. \textbf{63}, 224211 (2020).
\bibitem{Yang2020} Z. B. Yang \textit{et al.}, Ann. Phys. (Berlin) \textbf{532}, 2000196 (2020).
\bibitem{Collett84} Collett M J and Gardiner C W 1984 Phys. Rev. A 30 1386
\bibitem{Agarwal06} Agarwal G S 2006 Phys. Rev. Lett. 97 023601
\bibitem{He07} He W P and Li F L 2007 Phys. Rev. A 76 012328
\bibitem{Yan12} Yan Z H, Jia X J, Su X L, Duan Z Y, Xie C D and Peng K C 2012 Phys. Rev. A 85 040305
\bibitem{Chen09} Chen H X and Zhang J 2009 Phys. Rev. A 79 063826 
\bibitem{Shang10} Shang Y N, Jia X J, Shen Y M, Xie C D and Peng K C 2010 Opt. Lett. 35 853
\bibitem{Hou2025} R. Hou, W. Zhang, X. Han, H. F. Wang and S. Zhang. Scientific Reports, 15(1) (2025) 5145
\bibitem{Zhou15} Zhou Y Y, Jia X J, Li F, Yu J, Xie C D and Peng K C 2015 Scientific Reports 5 11132 
\bibitem{Malykin2000} G. B. Malykin, Phys. Usp. 2000, 43, 1229.
\bibitem{Maayani18} Maayani, Shai, Dahan, Raphael, Kligerman, Yuri, Moses, Eduard, Hassan, and Absar, Nature 558, 569 (2018)
\bibitem{Liang24} Z. Liang, Y. Wu, J. Li, Opt. Lett. 2024, 49, 2749.
\bibitem{Shen22} R. C. Shen, J. Li, Z. Y. Fan, Y. P. Wang, and J. Q. You. Physical Review Letters, 129(12), 123601 (2022).
\bibitem{LiuOL19} Z.-X. Liu, C. You, B. Wang, H. Xiong, and Y. Wu. Opt. Lett. 44, 507 (2019).
\bibitem{LiuIEEE19} Z.-X. Liu, H. Xiong, and Y. Wu. IEEE Access 7,
57047 (2019).
\end{thebibliography}
\end{document}